\documentclass[aps,superscriptaddress,floatfix,onecolumn,notitlepage]{revtex4-2}
\usepackage[a4paper,top=2.0cm,bottom=2.0cm,left=2.0cm,right=2.0cm]{geometry}
\usepackage{physics}
\usepackage{mathbbol}
\usepackage{amsmath}
\usepackage{amssymb}
\DeclareSymbolFontAlphabet{\amsmathbb}{AMSb}
\usepackage{feynmp-auto}
\usepackage{slashed}
\usepackage{dcolumn}
\usepackage{bm}
\usepackage{bbm}
\usepackage{color}
\usepackage{extarrows}
\usepackage{xfrac}
\usepackage[colorlinks=true]{hyperref}
\usepackage[x11names]{xcolor}
\usepackage{graphicx}
\graphicspath{{./}}
\definecolor{aeroblue}{rgb}{0.79,1.00,0.90}

\def\cdbox#1{\colorbox{aeroblue}{$\displaystyle{#1}$}}
\definecolor{black}{rgb}{0,0,0}
\newcommand{\red}{\textcolor{red}}

\newcommand{\He}{$^3$He}
\newcommand{\vv}{\vb{v}}
\newcommand{\vk}{\vb{k}}
\renewcommand{\vr}{\vb{r}}
\newcommand{\khat}{\hat\vk}
\def\sgn#1{\mathsf{sgn}(#1)}
\def\Tr#1{\mbox{Tr}\Big\{#1\Big\}}
\def\Trfour#1{\mbox{Tr}_4\Big\{#1\Big\}}

\def\cN{{\mathcal N}}
\def\cP{{\mathcal P}}
\def\sP{{\mathsf P}}
\renewcommand{\Im}{\mbox{Im\,}}

\newcommand{\mfD}{\mathfrak{D}}
\newcommand{\mfG}{\mathfrak{G}}
\newcommand{\mfF}{\mathfrak{F}}
\newcommand{\whmfN}{\widehat{\mathfrak{N}}}

\newcommand{\whmfG}{\widehat{\mathfrak{G}}}
\newcommand{\whmfGR}{\widehat{\mathfrak{G}}^{\mathrm{R}}}
\newcommand{\whDel}{\widehat{\Delta}}
\newcommand{\whSigR}{\widehat{\Sigma}^{\mathrm{R}}}
\newcommand{\whSig}{\widehat{\Sigma}}
\newcommand{\whtauz}{{\widehat{\tau}_3}}
\newcommand{\whtauy}{{\widehat{\tau}_2}}
\newcommand{\whtaux}{{\widehat{\tau}_1}}

\newcommand{\nf}{n_{\mathrm{F}}}

\def\ns{\negthickspace}
\def\point#1#2{{\sf #1}_{\mbox{\tiny #2}}}
\begin{document}
\title{Theory of Two-level Tunneling Systems in Superconductors}
\author{Junguang He}
\affiliation{Department of Physics and Astronomy, Northwestern University, 
             Evanston, IL 60208, USA}
\author{Wei-Ting Lin}
\affiliation{Hearne Institute of Theoretical Physics, 
             Louisiana State University, Baton Rouge, LA 70808, USA}
\author{J.~A. Sauls}
\email{sauls@lsu.edu}
\affiliation{Hearne Institute of Theoretical Physics, 
             Louisiana State University, Baton Rouge, LA 70808, USA}
\date{\today}
\begin{abstract}
We develop a field theory formulation for the interaction of an ensemble of two-level tunneling systems (TLS) with the electronic states of a superconductor.
Predictions for the impact of two-level tunneling systems on superconductivity are presented, including $T_c$ and the spectrum of quasiparticle states for conventional BCS superconductors. 
We show that non-magnetic TLS impurities in conventional s-wave superconductors can act as pair-breaking or pair-enhancing defects depending on the level population of the distribution of TLS impurities. 
We present calculations of the enhancement of superconductivity, both $T_c$ and the order parameter, for TLS defects in thermal equilibrium with the electrons and lattice.
The scattering of quasiparticles by TLS impurities leads to sub-gap states below the bulk excitation gap, $\Delta$, as well as resonances in the continuum above $\Delta$.
The energies and spectral weights of these states depend on the distribution of tunnel splittings, while the spectral weights are particularly sensitive to the level occupation of the TLS impurities. 
Under microwave excitation, or decoupling from the thermal bath, a nonequilibrium level population of the TLS distribution generates subgap quasiparticle states near the Fermi level which contribute to dissipation and thus degrade the performance of superconducting devices at low temperatures.
\end{abstract}

\maketitle

\section{Introduction}\label{sec1}

One of the key sources of decoherence that limits the performance of superconducting devices for quantum computing and quantum sensing applications is two-level tunneling systems (TLS) in the oxides that coat the surfaces and interfaces of superconducting niobium and aluminum~\cite{mcr20,abd22}.
Nb$_2$O$_5$ is known to limit the lifetime of microwave photons in superconducting RF cavities. Removal and prevention of re-growth of Nb$_2$O$_5$ leads to microwave photon lifetimes of $T_1\simeq 2\,\mbox{secs}$ in Nb SRF cavity resonators~\cite{rom20}.
Similarly, it was recently shown that removal of Nb$_2$O$_5$ from the Nb components of transmon quibits lead to significantly improved the qubit coherence times with $T_1\gtrsim 500\,\mu\mbox{sec}$~\cite{bal24}.
While removing surface oxide leads to improvement in microwave photon lifetimes in high-Q Nb SRF cavities and transmon coherence times, TLS disorder \emph{embedded in} bulk Nb likely remains and can limit further improvement in the quality factor of Nb SRF cavities or coherence times of superconducting devices. Evidence of residual TLS disorder is inferred from the suppression of the quality factor at very low temperatures \emph{after} the removal of Nb$_2$O$_5$ by heat treatment~\cite{abo25}.

Known sources of TLS defects embedded in Nb are low concentrations of hydrogen (H), deuterium (D) and oxygen (O), that diffuse into Nb. Oxygen traps hydrogen and deuterium in the Nb unit cell. H and D tunnel between two symmetry related tetrahedral sites that are the local minima of a double-well potential energy profile for H/D trapped by O~\cite{abo25}. Evidence of atomic tunneling of H and D atoms in Nb is provided by heat capacity~\cite{wip84}, ultra-sound~\cite{bel85}, and inelastic neutron scattering~\cite{mag83,mag86} measurements. Strain induced by the random spatial distribution of O-H and O-D impurities leads to a distribution of H/D tunnel splittings accessible by microwave photons~\cite{wip84,abo25}.

An additional source of decoherence for superconducting qubits is dissipation from quasiparticles under the nonequilibrium conditions of device operation. 
Losses from the \emph{generation} of nonequilibrium quasiparticles are reported to be as important as other loss channels in transmon qubits when coherence times reach $200\,\mu\mbox{sec}$~\cite{ser18}.
The effects of \emph{static} disorder - magnetic and non-magnetic - on superconductivity have been studied extensively by many authors~\cite{and59,abr59b,abr61,mak63} since the development of the microscopic theory of superconductivity~\cite{bar57}. For recent developments related to microwave circuits and SRF cavities see Refs.~\cite{sau22,uek22,uek23,zar23,uek24}.
There are a few studies of the effects of TLS impurities on superconducvity, including how the TLS impurities might mediate superconductivity~\cite{rie80}, or enhance $T_c$ in metallic glasses~\cite{bra70,har83}, or act as Kondo-type defects~\cite{mae84}.
In the opposite limit of isolated TLS impurities embedded in a superconductor, Yu and Granato considered the effect of pairing correlations on the magnitude of the tunnel splitting of an individual TLS, specifically a Hydrogen atom embedded in niobium~\cite{yu85}.
They found that the tunnel splitting increases by $\approx 200-700\,\mbox{Hz}$ in the superconducting state compared to that in the normal state of Nb. This increase in tunnel splitting for a single H TLS impurity is very small compared to the width in the distribution of H tunnel splittings resulting from the strain induced by O impurities that trap H in Nb, even for the lowest OH concentrations ($c_{\text{O}}=0.1\%$) measured~\cite{wip84}.

In this report we develop the theory of embedded TLS impurities in conventional superconductors, such as Nb and Al, which in the clean limit are fully gapped BCS superconductors. We consider the regimes of weak to moderate disorder with $\hbar/2\pi\tau T_c \lesssim 1$ relevant to superconducting devices for qubits and sensors. Thus, we are neither in the regime of metallic glasses, nor the ultra-clean limit of isolated TLS impurities.

Our focus is on the effects of TLS disorder on the superconducting transition and the excitation spectrum. Unlike static nonmagnetic impurites, TLS disorder is a random distribution of dynamical impurities, which can be pair-breaking or pair-enhancing depending on the distribution of tunnel splittings and the level occupancies of the TLS impurities.
One result of this study is that under equilibrium conditions we find a modest enhancement of $T_c$ and the excitation gap, qualitatively consistent with earlier work on the effect of TLS in metallic glass superconductors~\cite{bra70,har83}.
Another result of this study is that embedded TLS impurities generate sub-gap states below the clean limit excitation gap, $\Delta$, as well as resonances in the continuum above $\Delta$. The energies and spectral weights of these states depend on the distribution of tunnel splittings, with the spectral weights being particularly sensitive to the level occupations of the TLS impurities.
Under the application of microwave power the level population of the TLS distribution can be driven out of equilibrium, in which case quasiparticle-TLS interactions are predominantly pair-breaking and generate a substantial density of subgap quasiparticle states. In particular, we show that \emph{gapless superconductivity} can develop, and thus generate a substantial nonequilibrium population of quasiparticles which are a new channel for dissipation.

The rest of the paper is organized as follows: Sec.~\ref{sec-Model} describes the basic theoretical model defined by the Hamiltonian for a conventional BCS superconductor coupled to TLS impurities. We introduce Abrikosov's Fermion representation for the isospin operators describing each TLS impurity in Sec.~\ref{sec-TLS_Isospin}, then discuss Popov and Fedotov's method to eliminate unphysical states generated by the Fermion operators.
The Fermion representation and Popov-Fedotov projection are central to the theoretical analysis that follows. In Sec.~\ref{sec-TLS_Isospin} we develop perturbation theory for the corrections to the Gorkov propagator generated by the quasiparticle-TLS interaction. 
 
The distribution of TLS impurities is described by random variables for the position, orientation and tunnel splittings. Configuration averaging over the random distribution of TLS impurities is described in Sec.~\ref{sec-configuration_averages}, and defines the TLS impurity self energies for quasiparticles and Cooper pairs discussed in Sec.~\ref{sec-Self_Energy}. The functional form of these self energies are key results that allow us to calculate the effects of a random distribution of TLS impurities on the superconducting transition, order parameter and quasiparticle density of states (DOS).
In Sec.~\ref{sec-Tc_Gap_Enhancement} we report results for impact of TLS impurities on the equilibrium state of superconductivity, including the enhancement of $T_c$ and the superconducting order parameter. We also report results for the density of states and spectra of sub-gap and above-gap resonance states resulting from multiple scattering by TLS impurities with equilibrium population levels.
In Sec.~\ref{sec-Nonequilibrium_TLS} we investigate the impact of TLS impurities with level occupations that are out-of-equilibrium with respect to the thermal bath of electron and phonon excitations. A key observation is that non-equilibrium TLS impurities are strong pair breakers generating a substantial density of low-energy sub-gap states that are likely sources of dissipation for disordered superconducting devices under microwave excitation.

\section{Theoretical Model}\label{sec-Model}

The modern theory of superconductivity is based on a finite-temperature quantum field theoretical formulation for both thermodynamics and nonequilibrium transport and dynamics. The dynamics of TLS impurities is based on an isospin representation for the quantum state of the TLS impurities and their coupling to the electronic degrees of freedom. 
However, the isospin operators describing the dynamics of the TLS impurities do not obey the Wick theorem that is central to the traditional quantum field theory formalism, specifically the diagrammatic formalism used to define and classify the terms contributing to the electronic self energy resulting from the interaction with TLS impurities.
We adopt the Fermion method introduced by Abrikosov~\cite{abr65} that allows us to represent the isospin operators in terms of local Fermion fields for the states of each TLS impurity, and the projection method of Popov and Fedotov~\cite{pop88} to eliminate unphysical states of the local Fermions. The representation of TLS impurities in terms of localized Fermions allows us to take advantage of standard quantum field theory methods to calculate the effects of TLS impurities on pairing correlations, $T_c$, and the quasiparticle spectrum of superconductors, as well as nonequilibrium properties such as the response to electromagnetic fields.  

We introduce the Hamiltonian for a BCS superconductor coupled to a distribution of TLS impurity atoms whose local dynamics is defined by tunneling in a double-well potential. We start from the Hamiltonian 
\begin{equation}
H= H_{\text{BCS}} + H_\text{TLS} + H_\text{e-TLS}
\,,\quad\mbox{where}
\label{eq-Hamiltonian}
\end{equation} 
\begin{eqnarray}
H_{\text{BCS}}
=
\int\ns d^3r
\psi^{\dag}_{\alpha}(\vb{r})\left(-\frac{\hbar^2}{2m}\nabla^2-\mu\right)\psi_{\alpha}(\vb{r})
-
\int\ns d^3r
\left\{
\psi^{\dag}_{\alpha}(\vb{r})
\Delta_{\alpha\beta}(\vb{r})
\psi^{\dag}_{\beta}(\vb{r})
+
\psi_{\alpha}(\vb{r})
\Delta^{\dag}_{\alpha\beta}(\vb{r})\,
\psi_{\beta}(\vb{r})
\right\}
,
\end{eqnarray} 
is the mean-field BCS Hamiltonian defined in terms of the field operators, $\psi_{\alpha}(\vr)$, for the spin $s=\frac{1}{2}$ conduction electrons, and an attractive interaction for spin-singlet, s-wave pairing, $V_{\alpha\beta:\gamma\rho}(\vr,\vr')=-\frac{1}{4}\,g\,\delta(\vr-\vr')\,(is_y)_{\alpha\beta}(is_y)_{\gamma\rho}$, where $s_y$ is the anti-symmetry Pauli spin matrix, 
and the $g>0$ is the pairing interaction strength. 
The mean-field pairing energy is then $\Delta_{\alpha\beta}(\vr)\equiv\Delta(\vr)(is_y)_{\alpha\beta}$, with $\Delta(\vr)=\frac{1}{2}\,g\,(-is_y)_{\alpha\beta}\langle\psi_{\alpha}(\vr)\psi_{\beta}(\vr)\rangle$,
where the condensate amplitude, $\langle\psi_{\alpha}(\vr)\psi_{\beta}(\vr)\rangle$, is the equal-time limit of Gorkov's anomalous propagator.

\begin{figure}
\includegraphics[width=0.5\linewidth]{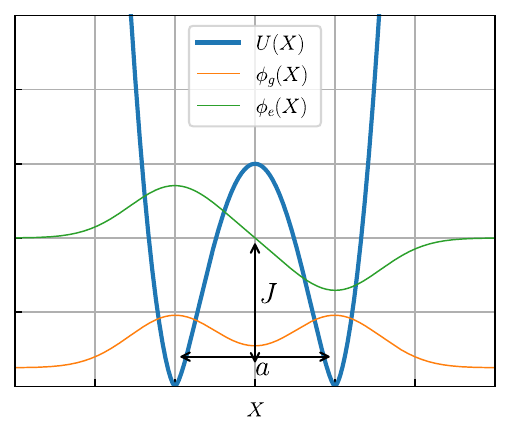}
\begin{minipage}{0.65\linewidth}
\caption{Double well potential with the ground and excited state wave functions shown for the tunneling impurity. The tunnel splitting is $J$, and the distance between the two potential energy minima is $a$.
\label{fig-DoubleWell}
}
\end{minipage}
\end{figure}

The second term in Eq.~\eqref{eq-Hamiltonian} is the Hamiltonian for the dynamics of $N$ TLS impurities of mass $M$ in a double-well potential, $U(\vb{X})$, such as that in Fig.~\ref{fig-DoubleWell},
\begin{equation}
H_\text{TLS}
=
\sum_{j=1}^{N}
\left\{
\frac{|\vb{P}_j|^2}{2M}
+
U(\vb{X}_j)
\right\}
\,,
\label{eq-H_atom}
\end{equation}
where $\vb{X}_j$ and $\vb{P}_j$ are canonically conjugate coordinates and momenta for the tunneling atom at the site $\vb{R}_j$.
%
%
The dynamics of a single TLS impurity is governed by the Hamiltonian in Eq.~\eqref{eq-H_atom} for $N=1$, with the double-well potential, $U(\vb{X})$, of the form shown in Fig.~\ref{fig-DoubleWell}. The two lowest energy states are the result of atomic tunneling through the barrier of the double-well potential between otherwise degenerate local minima separated by a distance $a\sim$ \AA. These are the symmetric (ground), $\ket{g}$, and anti-symmetric (excited), $\ket{e}$ states with the wavefunctions, $\phi_{g,e}(X)$, as sketched in Fig.~\ref{fig-DoubleWell}.

In what follows we retain only the two low-lying tunnel-split states of each impurity, i.e. we assume the high-energy states of the impurity Hamiltonian in Eq.~\eqref{eq-H_atom}, which are separated from the ground doublet by $\delta E_{\text{imp}}$, are inaccessible at low temperatures, $k_{\mbox{\tiny B}}T\ll \delta E_{\text{imp}}$, and low microwave photon energies, $\hbar\omega\ll\delta E_{\text{imp}}$.
Thus, the Hamiltonian for the TLS impurities reduces to
\begin{equation}
H_\text{TLS}=\sum_j\frac{E_j}{2}\,\sigma_z(j)
\,,
\label{eq-H_TLS}
\end{equation}
where $E_j$ is the energy difference between the excited and ground states of the $j$-th impurity, and $\sigma_z(j)$ is the diagonal Pauli matrix for the isospin representation of the $j^{th}$ two-level tunneling atom.
For identical, non-interacting TLS impurities in a symmetric double-well potential the energy splittings $E_j$ are degenerate and equal to the tunnel splitting matrix element $J$.
However, the random distribution of TLS impurities generates a strain field within the metal that generates asymmetry in the minimum energies, $\{\varepsilon_j | j=1,\ldots, N\}$, of the double-well potential for each impurity, as well as a random distribution of tunnel matrix elements, $\{J_j| j=1,\ldots, N\}$, leading to a distribution of energy level splittings, 
\begin{equation}
E_j = \sqrt{J_j^2 + \varepsilon_j^2}
\,.
\end{equation}

The third term in Eq.~\eqref{eq-Hamiltonian} describes the interaction between the conduction electrons and the TLS impurities,
\begin{equation}
H_\text{e-TLS} 
= 
\sum_{\alpha}\int d^3r\,
\psi^{\dag}_{\alpha}(\vb{r})
\sum_{j}V(\vb{r}-\vb{X}_j)
\psi_{\alpha}(\vb{r})
\,,
\label{eq-e-TLS}
\end{equation}
where $V(\vb{r}-\vb{X}_j)$ is the interaction potential between an electron at $\vb{r}$ and the $j^{th}$ TLS impurity with position operator $\vb{X}_j$. If the impurity positions are static, e.g. $M\rightarrow\infty$, then, $H_\text{BCS}+H_\text{e-TLS}$ defines the Hamiltonian for superconductors with embedded static impurities. 
However, TLS impurities are dynamical, described by $H_\text{TLS}$ and the coupling to the electronic system, $H_{\text{e-TLS}}$. 
In general non-TLS impurities are also present in the superconductor, giving rise to an electron-impurity interaction that can be treated using standard T-matrix methods to describe the static random potential~\cite{rai94}. In the case of conventional isotropic BCS superconductors static non-magnetic impurities do not suppress $T_c$, modify the equilibrium gap or generate quasiparticle states below the gap~\cite{and59,abr59b}, but do limit the a.c. conductivity and field penetration into the superconductor, and can be included using standard methods~\cite{rai94}. In what follows we focus on the role of TLS impurities.
This model was considered by Maekawa et al.~\cite{mae84}. Our formulation, analysis and results differ significantly from theirs, and other authors as discussed in Sec.~\ref{sec-earlier_TLS-QP_research}. We report our results, with discussion, in Secs.~\ref{sec-TLS_Isospin} and \ref{sec-Tc_Gap_Enhancement}.~\footnote{Maekawa et al. consider the Kondo limit of zero tunnel splitting, while we consider the opposite limit of a distribution on tunnel splittings in which case no Kondo singularity appears in our results for $T_c$.}

The matrix elements for the interaction of electrons and a TLS impurity then reduce to 
\begin{equation}
V_{ba}(\vb{r})
=
\int\dd[3]X\,\phi_b^*(\vb{X})\,V(\vb{r}-\vb{X})\,\phi_a(\vb{X})
\,,
\end{equation}
where $a,b\in\{e,g\}$. 
Thus, the electron-TLS interaction potential, for a single TLS impurity, can be expressed as a $2\times 2$ matrix in the TLS isospin space,
\begin{eqnarray}\label{int_pot}
V(\vb{r})
&=&
\begin{pmatrix}
\mel{e}{V}{e} & \mel{e}{V}{g}
\cr
\mel{g}{V}{e} & \mel{g}{V}{g}
\end{pmatrix}
\,,
\end{eqnarray}
or in terms of the Pauli matrix representation~\footnote{We use $(s_x,s_y,s_z)$ to denote Pauli matrices in spin space, $(\sigma_x,\sigma_y,\sigma_z)$ to denote Pauli matrices in the TLS isospin space, and $(\whtaux,\whtauy,\whtauz)$ to denote Pauli matrices in the particle-hole (Nambu) space.}
\begin{equation}
V(\vb{r}) =
v(\vb{r})\,\mathbb{1}+ m(\vb{r})\,\sigma_z + n(\vb{r})\,\sigma_x
\,.
\end{equation}
The $\sigma_x$ term is the matrix element for electron scattering by the TLS impurity that is combined with an impurity transition from the ground (excited) to the excited (ground) state. The absence of a $\sigma_y$ term is because we chose the phases of the wave functions, $\phi_{e,g}(X)$, to be real. The $v$ term, proportional to the $2\times 2$ unit matrix, $\mathbb{1}$, corresponds to the matrix element for the interaction of electrons with a static impurity, i.e. neglecting the level splitting and transition amplitude of the TLS impurity.

Thus, the interaction of electrons with a distribution of TLS impurities is defined by the Hamiltonian, 

\begin{equation}
H_\text{e-TLS}
=
\sum_{\alpha}\int\dd[3]r\,\psi^\dagger_{\alpha}(\vb{r})
\sum_j
\bigg[
v_j(\vb{r})\,\mathbb{1}
+ 
m_j(\vb{r})\,\sigma_{z}(j)
+ 
n_j(\vb{r})\,\sigma_{x}(j)
\bigg]
\psi_{\alpha}(\vb{r})
\,,
\end{equation}
or upon transforming the field operators to the momentum-space representation,
\begin{equation}\label{H_TLS}
H_\text{e-TLS}
=
\sum_{\alpha}\frac{1}{V}\sum_{\vb{k}',\vb{k}}
c^{\dag}_{\vb{k}'\alpha}
\sum_j
\left[
v_{j}(\vb{k}',\vb{k})\,\mathbb{1}
+
m_{j}(\vb{k}',\vb{k})\,\sigma_{z}(j)
+
n_{j}(\vb{k}',\vb{k})\,\sigma_{x}(j)
\right]\,
c_{\vb{k}\alpha}
\,,
\end{equation}
where $f(\vb{k}',\vb{k})\equiv\int d^3r\,e^{-i(\vb{k}'-\vb{k})\cdot\vb{r}}\,f(\vb{r})$ define the matrix elements for the electron-TLS interactions in Eq.~\eqref{H_TLS}.
Transformation to the momentum space representation also reduces the mean field BCS Hamiltonian to the standard form for a conventional isotropic superconductor,
\begin{equation}
\label{eq-BCS_mf-Nambu_momentum_space}
H_\text{BCS}=\sum_{\vb{k}}
\Psi^\dagger_{\vb{k}}
\Big(\xi_{\vb{k}}\,\widehat{\tau}_3
+
\Delta\,
(is_y)(-i\widehat{\tau}_2)
\Big)
\Psi_{\vb{k}},
\end{equation}
where 
$\Psi^\dagger_{\vb{k}}=(c^{\dag}_{\vb{k}\uparrow}\,,c^{\dag}_{\vb{k}\downarrow}\,,c_{-\vb{k}\uparrow}\,,c_{-\vb{k}\downarrow})$ is the Nambu spinor for the creation (annihilation) of an electron of momentum $\vb{k}$ ($-\vb{k}$) and spin $\uparrow$ or $\downarrow$, $\xi_{\vb{k}}=k^2/2m-\mu$ is the kinetic energy of electrons with momentum $\pm\vb{k}$ relative to the chemical potential, and $\widehat\tau_{1,2,3}$ denote the $2\times 2$ Pauli matrices in particle-hole space. Note that we have chosen the global phase of the order parameter $\Delta$ to be real and positive.

\section{TLS Isospin, Fermion Representation \& Statistical Mechanics}\label{sec-TLS_Isospin}

The isospin operators describing the dynamics of the TLS impurities do not obey the Wick theorem for a perturbation expansion in the framework of quantum field theory. 
We can circumvent this limitation by introducing local Fermion operators \`a la Abrikosov that create the ground and excited state of the TLS, i.e. $f_{a}^{\dag}\ket{0}=\ket{a}$ where $a\in\{e,g\}$ corresponds to the two levels of the TLS impurity~\cite{abr65}.

\begin{figure}
\begin{minipage}{0.5\linewidth}
\begin{center}
\includegraphics[width=0.70\linewidth]{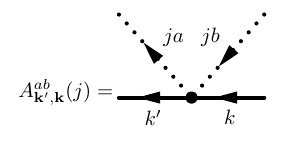}
\caption{Feynman diagram for the interaction vertex of conduction electrons and pseudo Fermions.
\label{fig-electron-pseudo-Fermion}}
\end{center}
\end{minipage}
\end{figure}

The dynamics of the TLS isospin operator is then encoded in the local Fermions, ``pseudo Fermions'' hereafter,
\begin{equation}\label{eq-pseudo_Fermion}
\va{\sigma}(j)=\sum_{a,b}f^{\dag}_{j,a}\,\va{\sigma}_{ab}\,f_{j,b}
\,,
\end{equation}
where $\va{\sigma}_{ab}$ are the matrix elements of the isospin Pauli matrices. The electron-TLS interaction Hamiltonian can then be written as
\begin{eqnarray}
\label{eq-H_electron_pseudo_Fermion}
H_{\text{e-TLS}} 
= 
\frac{1}{V}\sum_{\vb{k}',\vb{k};\alpha}\sum_{j}
c^\dagger_{\vb{k}'\alpha}\,v_j(\vb{k}',\vb{k})\,c_{\vb{k}\alpha}
+ 
\frac{1}{V}\sum_{\vb{k}',\vb{k};\alpha}\sum_{j;a,b}
c^\dagger_{\vb{k}'\alpha}\,f^{\dag}_{j,a}\,A_{\vb{k}',\vb{k}}^{ab}(j)\,f_{j,b}\,c_{\vb{k}\alpha}
\end{eqnarray}
where $v_{j}(\vb{k}',\vb{k})$ is the static contribution to the matrix element for quasiparticle scattering by the random potential of TLS impurities, and 
\begin{equation}
A_{\vb{k}',\vb{k}}^{ab}(j)
=
m_{j}(\vb{k}',\vb{k})\,\left(\sigma_z\right)_{ab} 
+ 
n_{j}(\vb{k}',\vb{k})\,\left(\sigma_x\right)_{ab}
\,,
\end{equation}
are the amplitudes that depend on the internal state of the TLS atoms encoded by the pseudo-Fermion operators and the Pauli matrix elements for each TLS impurity located at the random sites labelled by index $j$. This vertex is represented by the Feynman diagram in Fig.~\ref{fig-electron-pseudo-Fermion}.
However, the transformation defined by Eq.~\eqref{eq-pseudo_Fermion} introduces unphysical states, i.e. the empty state $\ket{0}$ and the doubly occupied state $\ket{eg}$, both of which must be excluded in carrying out ensemble averages of physical operators~\cite{abr65}.

\subsection{Popov-Fedotov Projection}\label{sec-PF}

The Nambu Green's function in the Matsubara formalism for a superconductor with an embedded distribution of TLS impurities is given by
\begin{equation}
\label{eq-Gorkov_Propagator}
\widehat{G}(x_1,x_2)=-\ev{T_\tau\Psi(x_1)\Psi^\dagger(x_2)}
\,,
\end{equation}
where $\Psi(x)=\mbox{col}\left(\psi_{\uparrow}(\vb{r})\,,\psi_{\downarrow}(\vb{r})\,,\psi_{\uparrow}^{\dag}(\vb{r})\,,\psi_{\downarrow}^{\dag}(\vb{r})\right)$ is the $4$-component Fermion Nambu spinor, and the ensemble average is taken over the grand canonical ensemble defined by the Hamiltonian,
\begin{equation}
H = H_{\text{BCS}} + H_{\text{TLS}} + H_{\text{e-TLS}}
\,,
\end{equation}
with Eqs.~\eqref{eq-BCS_mf-Nambu_momentum_space},~\eqref{eq-H_TLS}, and~\eqref{eq-H_electron_pseudo_Fermion}.
The average denoted by $\langle\ldots\rangle$ in Eq.~\eqref{eq-Gorkov_Propagator} is a constrained ensemble average over pseudo Fermion states with occupancy $n_f=1$ corresponding to the two physical states $\ket{e}$ and $\ket{g}$ for any of the TLS impurities,
\begin{equation}
\label{eq-constrained_ensemble_average}
\ev{\mathcal O}\equiv e^{\beta\Omega}\Tr{e^{-\beta H}\mathcal O}\eval_{n_f=1}
\,.
\end{equation}
The partition function, and corresponding thermodynamic potential, are also calculated in the physical subspace,
\begin{equation}
e^{\beta\Omega}=1/\Tr{e^{-\beta H}}\eval_{n_f=1}
\,.
\end{equation}
As formulated the constraint is incompatible with finite-temperature quantum field theory where ensemble averages are performed over the entire Hilbert space. In order to circumvent this problem and still enforce the constraint we use the Popov-Fedotov method. 
Note that pseudo-Fermion number, $\sum_a f^{\dag}_a f_a$, is conserved by the Hamiltonian $H_{\text{e-TLS}}$. Thus, following Popov and Fedotov, we can implement the constrained statistical average as an unconstrained trace over states of conduction electrons and the pseudo-Fermion Fock space by introducing a pseudo-Fermion chemical potential,
\begin{equation}
\ev{\cdots}=e^{\beta\Omega}\Tr{e^{-\beta[H-\mu_f(n_f-1)]}\cdots}
\,.
\end{equation}
The physical constraint $n_f=1$ is achieved by choosing $\mu_f=-i\pi T/2$, which enforces zero statistical weight to the pseudo-Fermion Fock states with $n_f=0$ and $n_f=2$, thus eliminating the unphysical states from the ensemble average~\cite{pop88}.
Specifically, the terms introduced by the chemical potential $e^{\beta\mu_f(n_f-1)}$ have no effect on the trace over physical states $\ket{n_f=1}$, but introduce a factor of $+i(-i)$ for the unphysical states $\ket{n_f=0}$ ($\ket{n_f=2}$). These phase factors enforce cancellation of the contributions to the ensemble average from the Fock states with $n_f=0$ or $n_f=2$.
The key result is that Abrikosov's pseudo-Fermion representation of the TLS isospin with the Popov-Fedotov projection of the unphysical pseudo-Fermion states allows us to employ traditional quantum field theory techniques, including perturbation theory based on the Wick theorem and the corresponding Feynman rules.

\begin{figure}
\centering
\includegraphics[width=0.75\linewidth]{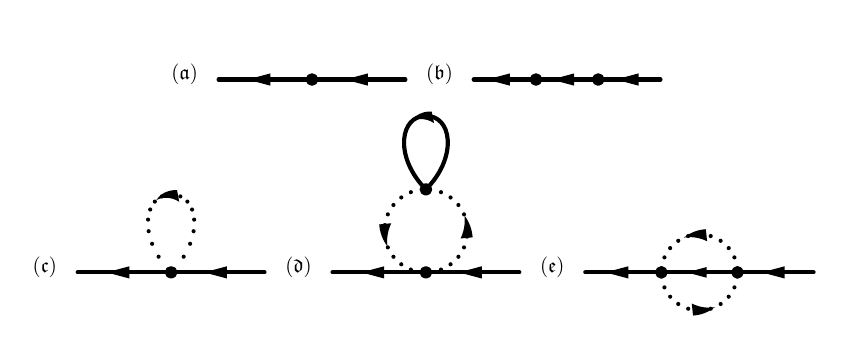}
\vspace*{7mm}
\begin{minipage}{0.75\linewidth}
\caption{
Leading order diagrams for the quasiparticle-TLS impurtiy scattering corrections to the Green's function. Solid lines represent the Gorkov propagator for quasiparticles and pairs, dotted lines represent the propagator for pseudo Fermions, solid circles represent the interaction vertex between Fermions and TLS impurity potential. Diagrams ($\mathfrak{a}$) and ($\mathfrak{b}$) represent the interaction with the static component of the random potential generated by the distribution of impurities. Diagram ($\mathfrak{c}$) is the 1$^{st}$ order contribution from pseudo-Fermions and the random potential, while diagrams ($\mathfrak{d}$) and ($\mathfrak{e}$) encode intermediate states of the TLS impurities. 
\label{fig-propagator_diagrams}
}
\end{minipage}
\end{figure}

For example, we can now calculate the unperturbed pseudo-Fermion propagator,
\begin{equation}
D_{i,a;j,b}(\tau-\tau')
\equiv
-\ev{T_\tau f_{i,a}(\tau)f^{\dag}_{j,b}(\tau')}_0
\,,
\end{equation}
where the subscript $i,a$ denotes the $a$-state of the pseudo-Fermion field operator of the $i^{th}$ impurity. In the Matsubara formalism we have
\begin{equation}
\label{f_greens_func}
D_{i,a;j,b}(\varepsilon_n)
=
\frac{\delta_{ij}\delta_{ab}}
     {i\varepsilon_n-\epsilon_{j,a}+\mu_f}
=
\frac{\delta_{ij}\delta_{ab}}
     {i\slashed{\varepsilon}_n-\epsilon_{j,a}}
\,,
\end{equation}
where $\slashed{\varepsilon}_n=\left(2n+\frac{1}{2}\right)\pi T$ is the shifted Matsubara energy, and $\epsilon_{j,\pm}=\pm E_j/2$ are the two energy levels of the $j^{th}$ TLS impurity. 

The presence of two-level tunneling centers enlarges the Hilbert space to include the quantum states of the random distribution of TLS impurities in the ensemble average. The statistical averaging is implemented by introducing pseudo-Fermions and the Popov-Fedotov constraint to represent the quantum states and dynamics of the TLS impurities as described above. In particular, we can formulate Feynman perturbation theory to calculate the corrections to the quasiparticle and Cooper pair propagators.
Figure~\ref{fig-propagator_diagrams} shows the Feynman diagrams for the corrections to the Nambu matrix propagator through second-order in the random potential of the TLS impurity distribtution.

The solid arrowed lines represent the Gorkov propagator for quasiparticles and pairs, while the dotted lines with arrows represent the propagator for pseudo Fermions. The solid circles represent the interaction vertex between Fermions/pairs and random potential. 
The first two diagrams, ($\mathfrak{a}$) and ($\mathfrak{b}$), represent the interaction with the \emph{static} component of the random TLS potential. The other three diagrams, ($\mathfrak{c}$), ($\mathfrak{d}$) and ($\mathfrak{e}$), describe intermediate quantum states, mediated by the pseudo-Fermions, of the random TLS potential.

It is important to note that Maekawa at el.~\cite{mae84} developed a field theory approach to the role of TLS disorder on superconductivity, adopting Abrikosov's method~\cite{abr65} to eliminate the fictitious pseudo-Fermion states $\ket{0}$ and $\ket{eg}$. However, this method fails to obey the linked-cluster theorem~\cite{zaw69}, and thus includes unphysical diagrams in the self energy.
Implementing the Popov-Fedotov projection method avoids this error.

\vspace*{-3mm}
\subsection{Configurational averages}\label{sec-configuration_averages}

As formulated the TLS potential corresponds to a particular realization of a distribution of TLS impurities, which are located at random positions in the otherwise crystalline superconductor. The orientation of the two-sites of double-well potential is also a random variable. Finally, the strain introduced by the random distribution of impurities induces a stochastic distribution of tunnel barriers, distances between local minima, asymmeties in the local minima, and thus a statistical distribution of tunnel splittings of the TLS impurites. This configurational randomness is in addition to the statistical distribution of states defined by the external thermal environment in contact with the superconductor.
Thus, we consider the interaction of electrons and pairs of electrons with a dilute random distribution of TLS impurities following methods introduced by Edwards~\cite{edw58}, Anderson~\cite{and59} and Abrikosov and Gorkov~\cite{abr59b}. In particular, we treat the positions of the TLS impurities, $\left\{\vb{R}_j|j\in 1,2,\ldots,N\right\}$, as random variables described by an uncorrelated joint probability distribution, $P(\{\vb{R}_j\})$. This allows us to carry out configurational averages of the perturbation expansion of the Gorkov Green's function in powers of $H_{\text{e-TLS}}$. The resulting configurational average results in coarse-grained translational invariance for configurational averages of all terms of the perturbation expansion for the Green's function. Thus,
\begin{equation}
\overline{\widehat{G}_{k,k'}(\{\vb{R}_j\})}
=
\prod_{j=1}^{N_{\text{TLS}}}\int\frac{d^3R_j}{V}\,
\widehat{G}_{k,k'}(\{\vb{R}_j\})
=
\delta_{\vb{k},\vb{k}'}\,\widehat{G}_{k}
\,.
\end{equation}
We generalize the conventional theory of dilute random impurities~\cite{edw58,and59,abr59b} to include configurational averaging over TLS impurities, including the orientation of the tunneling path, as well as the distribution of tunnel splittings. Averaging over positions of the TLS impurities follows the Edwards, Anderson and Abrikosov-Gorkov theory for a homogeneous distribution of uncorrelated randomly located impurities. 
The vertex for quasiparticle-impurity scattering depends on the random positions of the trapping potential of the TLS impurities, $\{R_j\}$, i.e. $A_{\vk',\vk}^{ab}(j)=e^{i(\vk'-\vk)\cdot\mathbf{R}_{j}}\tilde{A}_{\vk',\vk}^{ab}(j)$. Thus, after summing over random phase factors, $e^{i(\vk'-\vk)\cdot\mathbf{R}_{j}}$, translational invariance is recovered for the configurational averaged Green's function.
After configurational averaging over the random positions of TLS impurities the scattering amplitude still depends on the \emph{orientation} of the tunneling path, and thus the orientation of the tunneling path relative to the momentum $\vk$. Here we assume that the orientation of tunneling path of the TLS impurities is randomly distributed with a uniform probability distribution. As a result coarse-grained rotational invariance is recovered for the configurational averaged propagator. 

\begin{figure}
\hspace*{7mm}
\includegraphics[width=0.85\linewidth]{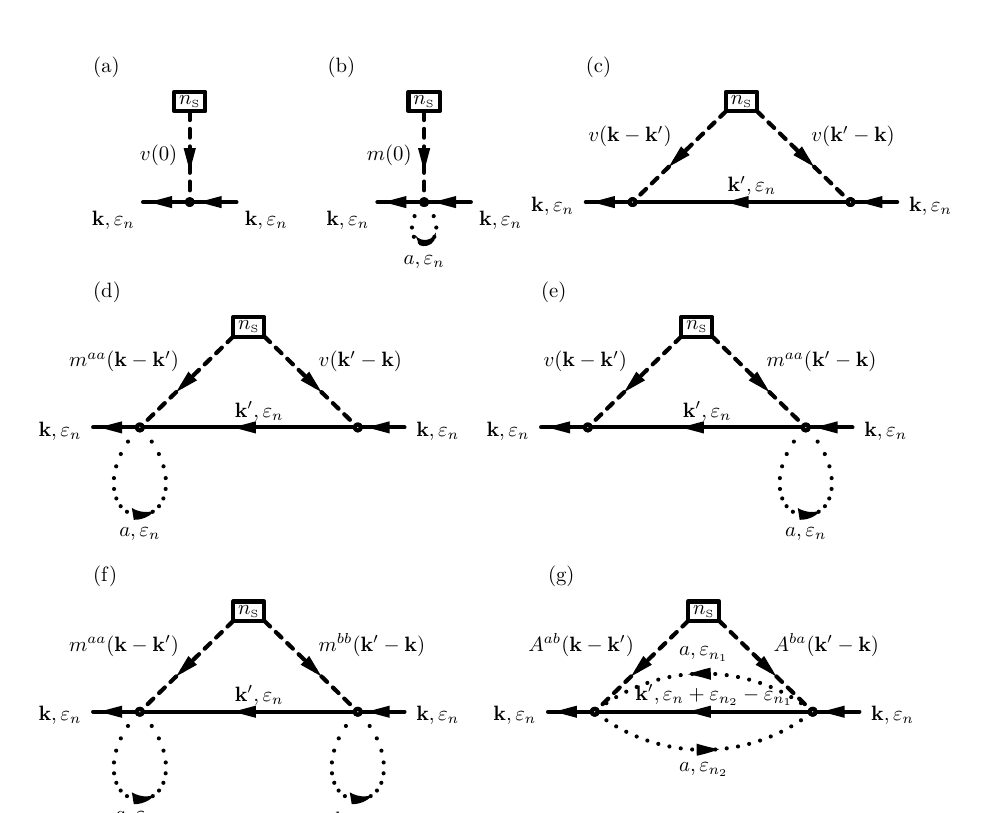}
\begin{minipage}{0.85\linewidth}
\vspace*{5mm}
\caption{
All contributions are proportional to the mean TLS density, $n_\text{S}=N_\text{S}/V$.
The top row are the 1$^{st}$ and 2$^{nd}$ order self energy diagrams from the static contribution to the TLS impurity scattering matrix element.
The second row are one-loop pseudo-Fermion corrections to the quasiparticle-TLS self-energy. These terms depend on both the static and dynamical interactions of the TLS impurities. 
The third row are two-loop pseudo-Fermion corrections to the quasiparticle-TLS self-energy. In particular, the last diagram describes multiple scattering by a TLS impurity including transitions between the excited and ground states of the TLS impurity.
\label{fig-configuration_averaged_diagrams}
}
\end{minipage}
\end{figure}

Configurational averaging the 1$^{st}$ and 2$^{nd}$ order terms in the perturbation expansion of the Nambu Green's function shown in Fig.~\ref{fig-propagator_diagrams} generates the set of Feynman diagrams shown in Fig.~\ref{fig-configuration_averaged_diagrams}.
These processes separate into two classes: those in which a quasiparticle exchanges energy with a TLS impurity, and those that do not.
Diagrams Fig.~\ref{fig-configuration_averaged_diagrams}(a) and Fig.~\ref{fig-configuration_averaged_diagrams}(b) are processes with no transfer of momentum or exchange of energy between the TLS impurities and quasiparticles. These terms simply renormalize the chemical potential.
Diagram Fig.~\ref{fig-configuration_averaged_diagrams}(c) is the 2$^{nd}$ order Born scattering process from the static contribution to the TLS impurity potential.
Diagrams Fig.~\ref{fig-configuration_averaged_diagrams}(d),(e) and (f), in which all pseudo-Fermion lines form loops connected at a single vertex, are quasiparticle self-energy terms that depend on the occupations of the TLS levels, but do not involve exchange of energy between quasiparticles and the TLS impurities.
The term shown in Fig.~\ref{fig-configuration_averaged_diagrams}(g) is the two-loop pseudo-Fermion correction to quasiparticle and Cooper pair self energy. In this process the pseudo-Fermion lines connect initial and final vertices, and thus describe transitions between the excited and ground states of the TLS. As a result energy is transferred between the TLS impurities and quasiparticles. This term plays the central role in pair-breaking and pair enhancement via quasiparticle-TLS scattering.

\section{Self Energies}\label{sec-Self_Energy}

The corrections to the Gorkov propagator for quasiparticles and pairs resulting from interactions between electrons and TLS impurities are encoded in the self energy functional via the Dyson equation,
\begin{equation}
\widehat{G}_k^{-1}
=
\widehat{G}^{(0)-1}_k - \widehat{\Sigma}_k
=
i\tilde{\varepsilon}_n\widehat{1}
-
\xi_{\vb{k}}\widehat{\tau}_3
-
\tilde{\Delta}(is_y)(-i\widehat{\tau}_2)
\,,
\end{equation}
where $\widehat{G}_k \equiv \widehat{G}(\vb{k},\varepsilon_n)$ and the self-energy can be expressed as $\widehat{\Sigma}_k=\Sigma_\varepsilon\widehat{1}+\Sigma_{\Delta}(is_y)(-i\widehat{\tau}_2)$ in Nambu space. The renormalized Matsubara energy and superconducting order parameter are
\begin{eqnarray}
i\Tilde{\varepsilon}_n &=& i\varepsilon_n - \Sigma_\varepsilon 
\,,
\\
\Tilde{\Delta} &=& \Delta + \Sigma_\Delta
\,.
\end{eqnarray}
The functional form of the contributions to the self energy functional follow from the Feynman diagrams. 

\subsection{Elastic scattering by TLS impurities}\label{sec-Self_Energy-elastic}

For the diagrams that do not involve energy exchange between quasiparticles and TLS impurities the terms can be expressed in terms of the various matrix elements, $v$, $m$, and $n$, and the pseudo-Fermion loops attached to the corresponding vertices.
For a pseudo-Fermion loop terms containing the $n$ matrix elements vanish because $\sigma_x$ is off-diagonal. Thus, the vertex pairs are (vv), (vm), (mv) and (mm), as shown in Figs.~\ref{fig-configuration_averaged_diagrams}(c),(d),(e) and (f). 
The matrix elements, $v_{\vk,\vk'},m_{\vk,\vk'},n_{\vk,\vk'}$, are slowly varying functions of the magnitude of the momenta, whereas the spectral weight of the propagator is concentrated near the Fermi level. Thus we can evaluate the matrix elements at $|\vk|=k_f$ and $|\vk'|=k_f$, in which case are functions of the directions, $\hat\vk$ and $\hat\vk'$, on the Fermi surface. Similarly, the normal-state density of states can be evaluated at the Fermi level, $N(\xi_{\vk})\rightarrow N(0)$. 
The corresponding self-energies reduce to,
\begin{eqnarray}
\widehat{\Sigma}^{(\text{vv})}_{k}
&=&
\frac{1}{V}\sum_j
N(0)
\ev**{v_{\vk,\vk'}v_{\vk',\vk}}_\Omega
\int\dd\xi_{\vk'}
\widehat{\tau}_3
\widehat{G}(\vk',i\varepsilon_n)
\widehat{\tau}_3,
\label{eq-Sigma-vv}
\\
\widehat{\Sigma}^{(\text{vm})}_{k} 
&=&
-2\frac{1}{V}\sum_j
N(0)
\sum_a
\ev**{v_{\vk,\vk'}m_{\vk',\vk}^{aa}}_\Omega
\,
T\sum_{n_1} 
D_{j,a}(\varepsilon_{n_1})
\int\dd\xi_{\vk'}\,
\widehat{\tau}_3
\widehat{G}(\vk',i\varepsilon_n)
\widehat{\tau}_3,
\label{eq-Sigma-vm}
\\
\widehat{\Sigma}^{(\text{mm})}_{k}
&=&
\frac{1}{V}\sum_j
N(0)
\sum_{ab}\ev**{m_{kk'}^{aa}m_{k'k}^{bb}}_\Omega
T\sum_{n_1} 
D_{j,a}(\varepsilon_{n_1})
T\sum_{n_2} 
D_{j,b}(\varepsilon_{n_2})
\,
\int\dd\xi_{\vk'}\,
\widehat{\tau}_3
\widehat{G}(\vk',i\varepsilon_n)
\widehat{\tau}_3,
\label{eq-Sigma-mm}
\end{eqnarray}
where $k\equiv(\vk,i\varepsilon_n)$, and $\langle\ldots\rangle_{\Omega}=\int\frac{d\Omega_{\khat}}{4\pi}\int\frac{d\Omega_{\khat'}}{4\pi}(\ldots)$ averages $\vk$ and $\vk'$ over the Fermi surface. The sign in Eq.~\eqref{eq-Sigma-vm} comes from the single pseudo-Fermion loop and the factor of $2$ is that diagrams (d) and (e) in Fig.~\ref{fig-configuration_averaged_diagrams} are equivalent after angular integration.

The pseudo-Fermion loops are evaluated using Cauchy's theorem to express the Matsubara sum as a contour integral of the Fermi function and the pseudo-Fermion propagator, analytically continued to the complex energy plane, enclosing the poles, $z_n=i\varepsilon_{n_1}$ of the Fermi function, $n_F(z)=1/(e^{\beta z}+1)$, then deforming the contour to encircle the pole of $D_{j,a}(z)$ at $z=\varepsilon_{j,a} - \mu_f$. The resulting residue gives the sum,
\begin{equation}
T\sum_{n_1} D_{j,a}(\varepsilon_{n_1}) = n_F(\epsilon_{j,a} -\mu_f) = 
\frac{1}{1 + i\,e^{\beta\epsilon_{j,a}}}
=
\frac{e^{-\beta\epsilon_{j,a}} - i}
     {e^{\beta\epsilon_{j}}+e^{-\beta\epsilon_{j}}}
\,.
\end{equation}
The denominator is the partition function for the two levels of the TLS $\epsilon_{j,g}=-\epsilon_j$ and $\epsilon_{j,e}=\epsilon_j > 0$, i.e. $Z_j = e^{\beta\epsilon_{j}}+e^{-\beta\epsilon_{j}}$, and thus, $N_{j,a} = e^{-\beta\epsilon_{j,a}}/Z_j$ is the equilibrium occupation of the two states of the $j^{th}$ TLS. Thus,
\begin{equation}
T\sum_{n_1} D_{j,a}(\varepsilon_{n_1}) = N_{a}(\epsilon_j) - i/Z(\epsilon_j)
\,.
\end{equation}
Using $m_{\vk',\vk}^{aa} = m_{\vk',\vk} \sigma_z^{aa}$ and carrying out the sum over level indices in Eqs.~\eqref{eq-Sigma-vm} and \eqref{eq-Sigma-mm}, we obtain,
\begin{equation}
\sum_a\,\sigma_z^{aa}\,T\sum_{n_1} D_{j,a}(\varepsilon_{n_1}) 
= 
N_{e}(\epsilon_j) - N_{g}(\epsilon_j) 
\,.
\end{equation}
Thus, the scattering terms in Eqs.~\eqref{eq-Sigma-vm} and~\eqref{eq-Sigma-mm} depend on the level populations of the states of the TLS, but these scattering processes do not change the level population of the TLS, and thus involve only elastic scattering of quasiparticles off the TLS impurities.
This is evident in that the Matsubara energy of the intermediate propagator in all of the above processes is $i\varepsilon_n$, i.e. for these scattering processes there is no energy exchange with the TLS impurity at either vertex.

\subsubsection{Quasiclassical Propagator and Self Energy}\label{sec-Eilenberger}

The Nambu matrix propagator appearing in Eqs.~\eqref{eq-Sigma-vv}-\eqref{eq-Sigma-mm} is integrated over the magnitude of the momentum $|\vk|$ near the Fermi surface, or equivalently the 
normal-state quasiparticle excitation energy, $\xi_{\vk}=v_f(|\vk|-k_f)$,
\begin{equation}
\whmfG(\hat\vk,\varepsilon_n)
\equiv
\int\dd\xi_{\vk}\,
\widehat{\tau}_3\widehat{G}(\vk,i\varepsilon_n)
\,.
\label{eq-quasiclassical_propagator}
\end{equation}
This is Eilenberger's quasiclassical propagator which is reduced to a function of position on the Fermi surface, $\hat\vk$. Note also the factor of $\whtauz$ as it is important in Eilenberger's reduction of the Dyson equation to a transport-like equation for the matrix $\whmfG(\hat\vk,\varepsilon_n;\vr)$ defined along trajectories specified by the normal-state group velocity, $\vv_{\vk}=v_f\hat\vk$, and locally for inhomogeneous equilibrium states,
\begin{equation}
\commutator{i\varepsilon_n\whtauz - \whDel(\hat\vk;\vr) - \whSig(\hat\vk,\varepsilon_n;\vr)}
           {\whmfG(\hat\vk,\varepsilon_n;\vr)}
+ 
i\vv_{\vk}\cdot\grad_{\vr}\whmfG(\hat\vk,\varepsilon_n;\vr) = 0
\,,
\label{eq-Eilenberger_Transport_Equation}
\end{equation}
where $\whDel(\hat\vk;\vr) = \Delta(\hat\vk;\vr)is_y\whtaux$ is the mean-field order parameter, which is defined here by the weak-coupling gap equation,
\begin{equation}
\Delta(\hat\vk;\vr) = \int \frac{d\Omega_{\vk'}}{4\pi}\,g(\hat\vk,\hat\vk')\,
T\sum_{\varepsilon_n}^{\omega_c}\,\mfF(\hat\vk',\varepsilon_n;\vr)
\,,
\end{equation}
where $g(\hat\vk,\hat\vk')$ is the pairing interaction within the bandwidth of attraction, $|\varepsilon_n|\le\omega_c$, and $\mfF(\hat\vk,\varepsilon_n;\vr)$ is Gorkov's anomalous propagator in the quasiclassical limit, i.e. the off-diagonal component of the matrix propagator, $\whmfG(\hat\vk,\varepsilon_n;\vr)$. 
The quasiclassical self energy is evaluated for momentum on the Fermi surface, and for the configurationally averaged TLS self energy defined by the diagrams in Fig.~\ref{fig-configuration_averaged_diagrams}, is post-multiplied by $\whtauz$, i.e. $\whSig(\hat\vk,\varepsilon_n)\equiv\whSig_k\whtauz$.
The Nambu matrix structure of the TLS self energies is determined by the Nambu matrix structure of the propagator, i.e. $\whmfG = \mfG\whtauz + \mfF\,is_y\whtaux$. Thus, we can express $\whSig = \Sigma_{\varepsilon}\whtauz + \Sigma_{\Delta}is_y\whtaux$.
The normalization of the propagator, which is absent in Eq.~\eqref{eq-Eilenberger_Transport_Equation}, is enforced by the Eilenberger's normalization condition~\cite{eil68},
\begin{equation}
\left[\whmfG(\hat\vk,\varepsilon_n)\right]^2 = -\pi^2\,\widehat{1}
\,.
\label{eq-Eilenberger_Normalization_Equation}
\end{equation}

In what follows we consider isotropic (``s-wave'') pairing of conduction electrons near an isotropic Fermi surface, i.e. $g(\hat\vk,\hat\vk')\rightarrow g$ and $\vv_{\hat\vk}=v_f\hat\vk$. Thus the mean field order parameter, propgator and self energy are all isotropic on the Fermi 
surface.
We also consider only homogeneous superconducting states in zero magnetic field, in which case the propagator and self energy are spatially uniform.

\subsection{Inelastic scattering by TLS impurities}\label{sec-SE-inelastic}

Diagram Fig.~\ref{fig-configuration_averaged_diagrams}(g) represents inelastic scattering of quasiparticles and pairs with the TLS impurities undergoing transitions between the ground and excited states. These processes can be pair-breaking or pair-enhancing depending on the distribution of tunnel splittings, the strength of the interaction and temperature. The self-energy for this process is derived below, and the effect of inelastic scattering by TLS impurities is described in the sections that follow. 

The expression for the inelastic self-energy obtained from diagram (g) in Fig.~\ref{fig-configuration_averaged_diagrams} is,
\begin{equation}
\whSig(\varepsilon_n)
=
-\frac{1}{V}\sum_{j}\,\sum_{ab}
N(0)\langle |A^{ab}_{\vk',\vk}(j)|^2 \rangle_{\Omega}
\,T\sum_{n_1}\,T\sum_{n_2}
\,D_{j,a}(\varepsilon_{n_1})D_{j,b}(\varepsilon_{n_2})
\,\whmfG(\varepsilon_{n}+\varepsilon_{n_2}-\varepsilon_{n_1})
\,,
\end{equation}
where we have integrated over $\xi_{\vk}$ to express the result in terms of the quasiclassical propagator, and we have post-multiplied by $\whtauz$. 
Configurational averaging over the positions of the TLS impurities follows the standard formulation by Edwards for a random distribution of dilute impurities in metals~\cite{edw58}.
The configurational average over random orientations of the tunneling atoms yields the Fermi surface average of the matrix elements, 
$\langle|A^{ab}|^2\rangle_{\Omega}
= \ev{|m_{\hat{\vk}'\hat{\vk}}|^2}_\Omega\,\delta^{ab} 
+ \ev{|n_{\hat{\vk}'\hat{\vk}}|^2}_\Omega\,\sigma_x^{ab}$.
The first term contributes to the elastic scattering rate, while the term proportional to $\sigma_x^{ab}$ defines the inelastic contribution the self-energy from quasiparticle-TLS scattering,
\begin{equation}
\Gamma^{ab}
\equiv 
n_s\,N(0)\,\langle|A^{ab}|^2\rangle_{\Omega}
=
\frac{\delta^{ab}}{2\pi{\tau_{el}}}
+\frac{\sigma_x^{ab}}{2\pi{\tau_{in}}}
\,,
\end{equation}
where $n_s = N_s/V$ is the density of TLS impurities, and 
$1/{\tau_{el}}\equiv 2\pi\,n_s\,N(0)\,
\ev{|m_{\hat{\vk}'\hat{\vk}}|^2}_\Omega$ 
and 
$1/{\tau_{in}}\equiv 2\pi\,n_s\,N(0)\,
\ev{|n_{\hat{\vk}'\hat{\vk}}|^2}_\Omega$ 
are the scattering rates for elastic and inelastic scattering by the TLS impurities, respectively. 
The tunnel splitting of each TLS impurity, $E_j$, is also a random variable 
which enters via the local Fermion propagators, $D_{j,a}(\varepsilon_n)$. We introduce the probability density, $p(E)$, for the distribution of tunnel splittings, and replace the sum over a particular realization of tunnel splittings by an ensemble average defined by this distribution, i.e.
$\frac{1}{N_s}\sum_j(\ldots) =\int dE\,p(E)(\ldots)$. In Sec.~\ref{sec-TLS_Distribution} we discuss several models for the distribution of tunnel splittings.
The resulting configuratonal averaged self energy functional is
\begin{equation}
\whSig(\varepsilon_n) = -\int dE\,p(E)\,
T\sum_{n_1}\,T\sum_{n_2}
\sum_{ab} 
\Gamma^{ab}\,
D_{a}(\varepsilon_{n_1}; E)
\,
D_{b}(\varepsilon_{n_2}; E)
\,
\whmfG(\varepsilon_{n}+\varepsilon_{n_2}-\varepsilon_{n_1})
\,,
\end{equation}
The Matsubara sum over the pair of local Fermion propagators defines a local Bosonic propagator with Matsubara frequency, $\omega_m\equiv\varepsilon_{n_1}-\varepsilon_{n_2}$, 
\begin{equation}
T\sum_{n_2}
D_{a}(\omega_m+\varepsilon_{n_2};E)
D_{b}(\varepsilon_{n_2}; E)
=
\frac{n_a-n_b}{\epsilon_a-\epsilon_b-i\omega_m}
\,,
\end{equation}
where $n_a=n_{\mathrm{F}}(\epsilon_a-\mu_f)$ and $\epsilon_{g/e}=\mp E/2$.
\begin{equation}
\whSig(\varepsilon_n)
=
-\int dE\,p(E)
\,T\sum_{m}
\,\sum_{ab}\Gamma^{ab}
\,\frac{n_a-n_b}{\epsilon_a-\epsilon_b-i\omega_m}
\,\whmfG(\varepsilon_{n}-\omega_m)
\,,
\end{equation}
Note that the $m=0$ term correponds to the elastic scattering, and thus does not contribute to the inelastic self energy functional. Summation over the level indices $a,b$ leads to
\begin{equation}
\whSig(\varepsilon_n)
=
\frac{1}{2\pi{\tau_{in}}}\,
\int dE\,p(E)\,N_{ge}(E)\,T\sum_{m\ne 0}\frac{2E}{\omega_m^2 + E^2}\,
\whmfG(\varepsilon_n-\omega_m)
\,,
\label{eq-Self_Energy-inelastic}
\end{equation}
where $N_g-N_e\equiv N_{ge}$ is the difference between the occupation of the ground level and the excited level of the TLS. For TLS impurities in equilibrium with the electrons and lattice at temperature $T$, $N_{ge}=\tanh(E/2T)$. Thus, for a distribution of tunnel splittings there can be a population of TLS impurities in their ground as well as a population of TLS impurities in their excited states.

The result for the inelastic TLS contribution to the self energy includes a sum over a Bosonic propagator, 
\begin{equation}
\mfD(\omega_m;E)
=\frac{2E}{\omega_m^2+E^2} 
=\left(\frac{1}{i\omega_m + E} - \frac{1}{i\omega_m - E}\right)
\,.
\label{eq-TLS_Bosonic_propagator}
\end{equation}
Indeed the self energy has a form similar to the self energy in Eliashberg's theory for phonon-mediated superconductivity. In the case of inelastic scattring by TLS impurities $\mfD(\omega_m)$ corresponds to the phonon propagator, the inelastic electron-TLS scattering rate, $1/2\pi\tau_{in}$, corresponds to the electron-phonon coupling, and the probability distribution of TLS tunnel splittings, $p(E)$, appears in place of the phonon density of states. 
In addition Eq.~\eqref{eq-Self_Energy-inelastic} includes the relative level population of TLS impurities, $N_{ge}(E)$.

The Bosonic TLS propagator originates from the pair of local Fermions that mediate transitions between the two levels of the TLS impurity. This leads to a retarded effective interaction that can enhance or suppress superconductivity depending on the distribution of TLS tunnel splittings and their relative occupations. 
It is also possible for the TLS Bosons to mediate superconductivity in the absence any other pairing mechanism. The effects of the TLS Bosons on superconductivity are discussed in detail in Secs.~\ref{sec-Tc_Gap_Enhancement} and ~\ref{sec-Nonequilibrium_TLS}. The self energy functional in Eq.~\eqref{eq-Self_Energy-inelastic} also depends on the propgator, $\whmfG(\varepsilon_n-\omega_m)$, which is also a function of the self energies for both quasiparticles and Cooper pairs. The complete set of equations must be solved self consistently as functions of the Matsubara energy, gap amplitude and temperature.

\subsection{TLS renormalized gap equation and $T_c$}

For homogeneous superconducting states Eilenberger's transport equation reduces to 
\begin{equation}\label{eq-Eilenberger_Equation-homogeneous}
\commutator{i\varepsilon_n\whtauz-\whDel-\whSig(\varepsilon_n)}{\whmfG(\varepsilon_n)}=0
\,.
\end{equation}
The solution for $\whmfG$ which satifies the normalization condition is then,
\begin{equation}
\whmfG(\varepsilon_n) 
= 
-\pi
\frac{(i\varepsilon_n-\Sigma_{\varepsilon})\whtauz-(\Delta+\Sigma_{\Delta})is_y\whtaux}
     {\sqrt{(\varepsilon_n+i\Sigma_{\varepsilon})^2 + (\Delta+\Sigma_{\Delta})^2}}
\,.
\label{eq-Renormalized_Propagator}
\end{equation}
Another observation is that the self energy terms corresponding to elastic scattering in Eqs.~\eqref{eq-Sigma-vv},\eqref{eq-Sigma-vm},\eqref{eq-Sigma-mm}, are all proportional to $\whmfG(\varepsilon_n)$ and thus drop out of the commutator in  Eq.~\eqref{eq-Eilenberger_Equation-homogeneous}. This is Anderson's theorem for non-magnetic static disorder in an isotropic conventional superconductor. Thus, only the inelastic contribution to the self energy from diagram Fig.~\ref{fig-configuration_averaged_diagrams}(g) contributes to the equilibrium propagator~\footnote{The elastic TLS self energy terms do contribute to nonequilibrium response functions that determine the microwave conductivity.}.

The impurity-renormalized propagator and impurity self energy must be solved self-consistently, a procedure that also includes the mean field order parameter determined by the gap equation and anomalous propagator,
\begin{equation}
\Delta = g\,\,\pi T\sum_{\varepsilon_n}^{\omega_c}\,
\frac{\Delta+\Sigma_{\Delta}}
     {\sqrt{(\varepsilon_n+i\Sigma_{\varepsilon})^2+(\Delta+\Sigma_{\Delta})^2}}
\,.
\label{eq-renormalized_gap_equation}
\end{equation}
The gap equation in the clean limit, 
\begin{equation}
\Delta = g\,\,\pi T\sum_{\varepsilon_n}^{\omega_c}\,
\frac{\Delta}{\sqrt{\varepsilon_n^2 + \Delta^2}}
\,,
\label{eq-gap_equation_clean-limit}
\end{equation}
is used to eliminate the pairing interaction and cutoff in favor of the clean limit transition temperature, $T_{c_0}$, and zero-temperature gap amplitude, $\Delta_0$. In particular, the linearized gap equation, 
$\Delta=g\,\pi T\sum_{\varepsilon_n}^{\omega_c}\,\Delta/|\varepsilon_n|
\equiv g\,K(T)\,\Delta$,
determines the transition temperature, $T_{c_0}$, for 
$\Delta\rightarrow 0^+$, i.e. $g\,K(T_{c_0})=1$, where in the limit 
$T\ll\omega_c$, $K(T)\equiv\pi T\sum_{\varepsilon_n}^{\omega_c}\,\frac{1}{|\varepsilon_n|}\simeq\ln\left(1.13\,\omega_c/T\right)$.
We then use $K(T)$ to regulate the log-divergent contribution to the sum in Eq.~\eqref{eq-renormalized_gap_equation} that is cutoff at $\omega_c$, and then use $g\,K(T_{c_0})=1$ to eliminate $g$ and $\omega_c$ in favor of $T_{c_0}$,
\begin{equation}
\Delta\,\ln\frac{T}{T_{c_0}} 
=
\pi T\sum_{\varepsilon_n}
\left(
\frac{\tilde\Delta}{\sqrt{\tilde\varepsilon_n^2+\tilde\Delta^2}}
-
\frac{\Delta}{|\varepsilon_n|}
\right)\,.
\label{eq-renormalized_gap_equation-regulated}
\end{equation}
The sum in Eq.~\eqref{eq-renormalized_gap_equation-regulated} converges on the scale set by $T_c$ and $\Delta$, and thus we can replace $\omega_c\rightarrow\infty$.
It is convenient to introduce the renormalized Matsubara frequency and gap parameter as
\begin{eqnarray}
i\tilde\varepsilon_n 
\equiv Z_{\varepsilon}\,i\varepsilon_n
= i\varepsilon_n-\Sigma_{\varepsilon}(\varepsilon_n)
\,,
\quad
\tilde\Delta 
\equiv Z_{\Delta}\,\Delta
= \Delta +\Sigma_{\Delta}(\varepsilon_n)
\,,
\label{eq-renormalization_functons}
\end{eqnarray}
where the renormalization functions are defined by the self energy functional in Eq.~\eqref{eq-Self_Energy-inelastic}, 
\begin{eqnarray}
Z_{\Delta}(\varepsilon_n)
&=& 
1 + 
\frac{1}{2\pi\tau_{in}}\,
\int dE\,p(E)\,N_{ge}(E)\,\pi T\sum_{\varepsilon_{n'}}\,
\mfD(\varepsilon_n-\varepsilon_{n'};E)\,
\frac{Z_{\Delta}(\varepsilon_{n'})}
     {\sqrt{\varepsilon_{n'}^2\,Z_{\varepsilon}(\varepsilon_{n'})^2 
           +\Delta^2\,Z_{\Delta}(\varepsilon_{n'})^2}}
\,,
\label{eq-Z_Delta}
\\
Z_{\varepsilon}(\varepsilon_n)
&=& 
1 + 
\frac{1}{2\pi\tau_{in}}\,
\int dE\,p(E)\,N_{ge}(E)\,\pi T\sum_{\varepsilon_{n'}}\,
\mfD(\varepsilon_n-\varepsilon_{n'};E)\,
\frac{Z_{\varepsilon}(\varepsilon_{n'})}
     {\sqrt{\varepsilon_{n'}^2\,Z_{\varepsilon}(\varepsilon_{n'})^2 
           +\Delta^2\,Z_{\Delta}(\varepsilon_{n'})^2}}
\,.
\label{eq-Z_varepsilon}
\end{eqnarray}

Before discussing the details of these results, we show the real-frequency expression of the self-energy, which allows us to calculate the spectrum of quasiparticle states, and to formulate a theory for pair-breaking by nonequilibrium TLS impurities.

\subsubsection{Analytic Continuation: Retarded Propagator and Self-Energy}\label{sec-real-freq}

In order to predict the effects of TLS impurities on the quasiparticle excitation spectrum we need to analytically continue the Matsubara propagator and self energy to the retarded functions defined on the real energy axis, i.e. $\whmfG(\varepsilon_n)\rightarrow\whmfGR(\varepsilon)$ 
and $\whSig(\varepsilon_n)\rightarrow\whSigR(\varepsilon)$ for $i\varepsilon_n\rightarrow\varepsilon+i0^+$. This is straightforward because the corresponding retarded and Matsubara functions are determined by the same spectral representation,
\begin{equation}
\whmfG(\varepsilon_n)
=
\int_{-\infty}^{+\infty}\ns\ns d\varepsilon'\,
\frac{\whmfN(\varepsilon')}{i\varepsilon_n - \varepsilon'}
\xlongrightarrow{i\varepsilon_n\rightarrow\varepsilon+i0^+}
\whmfGR(\varepsilon)
\equiv
\int_{-\infty}^{+\infty}\ns\ns d\varepsilon'\,
\frac{\whmfN(\varepsilon')}{\varepsilon+i0^+ - \varepsilon'}
=
\sP
\int_{-\infty}^{+\infty}\ns\ns d\varepsilon'\,
\frac{\whmfN(\varepsilon')}{\varepsilon - \varepsilon'}
-i\pi\whmfN(\varepsilon)
\,.
\label{eq-analytic_continuation}
\end{equation}
The first term on the right side of Eq.~\eqref{eq-analytic_continuation} is real and determined by the principal part integral over the real axis. The imaginary part is the spectral function which determines the quasiparticle density of states (DOS),
\begin{equation}
\cN(\varepsilon) = -\frac{N_f}{4\pi}\Im\Trfour{\whtauz\whmfGR(\varepsilon)} 
\,,
\label{eq-DOS}
\end{equation}
where the trace is over Nambu space.  
Similarly, the spectral function for the pairing amplitude is 
\begin{equation}
\cP(\varepsilon)=-\frac{N_f}{4\pi}\Im\Trfour{\whtaux(-is_y)\whmfGR(\varepsilon)}
\,.
\end{equation}
Analytic continuation of the Matsubara self energy to the real axis is carried out using the fact that the Fermi and Bose distribution functions have simple poles at the corresponding Matsubara frequencies.
Deforming contours surrounding the poles on the imaginary axis to the real axis we obtain the retarded self-energy,
\begin{eqnarray}
\widehat{\Sigma}^{\mathrm{R}}(\varepsilon) 
= 
\frac{1}{2\pi\tau_{in}} 
\int dE\,p(E)\, 
&\bigg\{&
\ns
\whmfG^{\mathrm{R}}(\varepsilon-E)\, N_g(E) 
+
\whmfG^{\mathrm{R}}(\varepsilon+E)\, N_e(E) 
\nonumber\\
&-&
\ns
N_{ge}(E)
\int_{-\infty}^{\infty}
\frac{d\varepsilon'}{2\pi i}
\frac{2E}{E^2-(\varepsilon+i0^+ - \varepsilon')^2}\,
2i\,\Im\whmfG^{\mathrm{R}}(\varepsilon')\,\nf(\varepsilon')
\bigg\} 
\,,
\label{eq-Self_energy-retarded}
\end{eqnarray}
where
\begin{equation}
\whmfG^{\mathrm{R}}(\varepsilon)
=
-\pi\frac{(\varepsilon-\Sigma^{\mathrm{R}}_\varepsilon(\varepsilon))\whtauz 
-(\Delta+\Sigma^{\mathrm{R}}_\Delta(\varepsilon))(is_y)\widehat{\tau}_1}
{\sqrt{(\Delta+\Sigma^{\mathrm{R}}_\Delta(\varepsilon))^2
-(\varepsilon-\Sigma^{\mathrm{R}}_\varepsilon(\varepsilon))^2}}
\,,
\label{eq-Propagator-retarded}
\end{equation}
is the retarded propagator. This expression for the retarded self energy defined as a function of excitation energy $\varepsilon$ relative to the Fermi level gives the same results for equilibrium properties as the Matsubara formulation. 
The advantage of the real energy formulation is that the spectral information, encoded in the retarded propagator, $\whmfGR(\varepsilon)$, is separated from the occupation information given by the Fermi distribution, $\nf(\varepsilon)$, as well as the TLS occupation factors, $N_g$, $N_e$, and $N_{ge}$, which in thermal equilibrium with electrons and phonons are determined by the Boltzmann factors, $e^{-E_{g,e}/T}$.
The real energy formulation allows us to investigate the effects of TLS impurities that are out of equilibrium relative to the electron-ion thermal bath, which we discuss in Sec.~\ref{sec-Nonequilibrium_TLS}.
Before reporting results we discuss models for the distribution of tunnel splittings of the TLS impurities. 

\subsubsection{Models for the Distribution of Tunnel Splittings}\label{sec-TLS_Distribution}

Even if all the tunneling atoms are identical, the local potential governing the tunneling of these atoms will vary in space due to a random distribution of impurities and defects. 
A random distribution of impurities introduces strain, and thus a random distribution of tunnel barriers, asymmetries in the local minima of the potential, as well as displacements of the local minima. Thus, the energy levels of the TLS impurities, particularly the tunnel splitting of the ground and excited state, which is exponentially sensitive to the tunnel barrier and displacement of the local minima, will form a distribution of tunnel splittings.
We introduce the probability distribution by replacing the sum over a specific configuration of all TLS impurities with an integration over a probability distribution of the tunnel splittings, i.e. 
\begin{equation}
\frac{1}{V}\sum_{j=1}^{N_s}\,F(E_j) 
\rightarrow
n_s\int dE\,p(E)\,F(E)
\,,
\end{equation}
where $n_s=N_s/V$ is the mean number density of TLS impurities and $p(E)$ is the probability density for a TLS impurity with tunnel splitting $E$.

\begin{figure}
\centering
\includegraphics[width=0.75\textwidth]{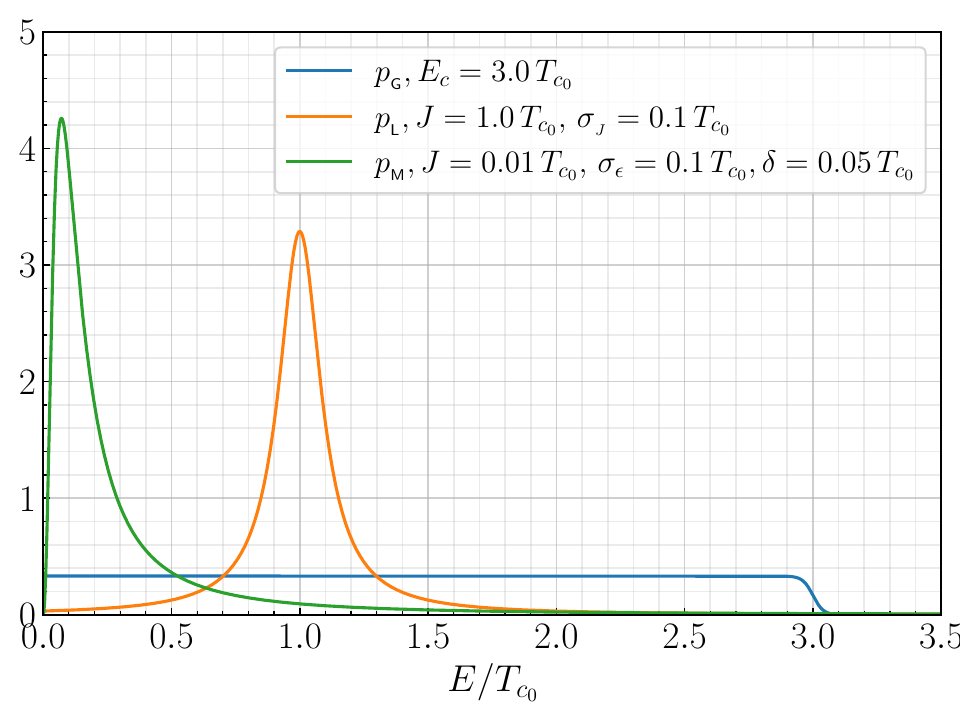}
\begin{minipage}{0.75\linewidth}
\caption{Three models for the distribution of TLS tunnel splittings used to predict the effects of TLS impurities on superconducting properties. The distributions, $p_{_\mathsf{G}}(E)$, $p_{_\mathsf{L}}(E)$ and $p_{_\mathsf{M}}(E)$, are defined in Sec.~\ref{sec-TLS_Distribution}.}
\label{fig-tls-dist}
\end{minipage}
\end{figure}

We consider three distributions as models for the TLS distribution of tunnel splittings:
(i) the distribution of tunnel splittings proposed for amorphous glasses~\cite{and72a,phi72}, 
(ii) a Lorentzian distribution with a spread in tunnel splittings, and 
(iii) a modified Lorentizian model introduced by Wipf and Neumaier for atomic tunneling in metals~\cite{wip84}.   
The glass model is defined by 
\begin{equation}
p_{_\mathsf{G}}(E) = \frac{1}{E_c}\Theta(E_c - E)
\,,
\end{equation}
where $E_c$ is a high energy cutoff. Thus, for $T\ll E_c$ the relevant density of tunnel splittings is constant and explains many aspects of atomic tunneling in amorphous glasses~\cite{and72a,phi72}. A key feature is there is substantial probability of TLS atoms with very low tunnel splitting.

For weaker disorder we consider a Lorentzian distribution of tunnel splittings based on the observation that small changes in the tunnel barrier or width of the double-well potential lead to significant changes in the tunnel splitting~\cite{abo25}. The Lorentizian distribution is defined by a most probable tunnel splitting, $J$, and a width of the distribution, $\sigma_{\ns_\mathit{J}}$,
\begin{equation}
p_{_\mathsf{L}}(E)=\frac{1}{\pi}\frac{\sigma_{\ns_\mathit{J}}}{(E-J)^2 + \sigma_{\ns_\mathit{J}}^2}
\,.
\end{equation}
Thus, $\sigma_{\ns_\mathit{J}}\ll J$ corresponds to the limit of weak strain disorder, while $\sigma_{\ns_\mathit{J}} \gg J$ corresponds to substantial strain disorder impacting the distribution of tunnel splittings.

We also consider a modified version of the distribution introduced by Wipf and Neumaier for H and D tunneling in Nb. These authors assumed that the random strain field leads to a distribution of asymmetry energies for the double well potential, but does not effect the tunnel matrix element, $J$. The resulting tunnel splitting for any particular tunneling atom is $E = \sqrt{J^2 + \epsilon^2}$, where $\epsilon$ is the asymmetry energy, i.e. the difference between the energies of the minima of the double well potential. Wipf and Neumaier assumed a Lorentzian distribution for the asymmetry energies with width $\sigma_{\epsilon}$, but assumed that the tunnel matrix element is the same for all TLS impurities. The resulting distribution is,
\begin{equation}
p_{_\mathsf{W}}(E)
=\frac{2}{\pi}
\frac{E}{\sqrt{E^2-J^2}}\,
\frac{\sigma_{\epsilon}}{E^2-J^2 + \sigma_{\epsilon}^2}\,
\Theta(E-J)
\,.
\end{equation}
This model worked well in accounting for the anomalous heat capacity of low concentrations of O, H and D in Nb~\cite{wip84}.
However, it has a square root singularity at a low-energy cut-off set by the tunnel matrix element $J$. We modify their distribution to account for variations in the tunnel matrix element due to strain disorder by introducing an imaginary part to remove the singularity and broaden the low-energy region associated with a distribution of tunnel matrix elements,
i.e.  
\begin{equation}
p_{_\mathsf{M}}(E)
=N_{_\mathsf{W}}\,
\Im\frac{E}{\sqrt{J^2 - (E+i\delta)^2}}\,
\frac{\sigma_{\epsilon}}{E^2-J^2 + \sigma_{\epsilon}^2}\,
\,,
\end{equation}
where $N_{_\mathsf{W}}$ is the prefactor that ensures that $p_{_\mathsf{M}}(E)$ is properly normalized which is computed numerically for any set of parameters, $J$, $\sigma_{\epsilon}$ and $\delta$.

The three distributions are shown in Fig.~\ref{fig-tls-dist} with the cutoff in the glass distribution, $p_{_\mathsf{G}}(E)$, set at $E_c = 3\,T_{c_0}$ corresponding to strong disorder with a broad distribution of tunnel splittings. For the Lorentzian distribution, $p_{_\mathsf{L}}(E)$, we set $J=T_{c_0}$ and $\sigma=0.1 T_{c_0}$, corresponding to relatively large tunnel splittings and weak disorder on the distribution of tunnel splittings. And for the modified Wipf-Neumaier distribution, $p_{_\mathsf{M}}(E)$, we set the peak in the distribution at relatively low energy, $J = 0.1\,T_{c_0}$, broadening $\delta = 0.01\,T_{c_0}$ and the spread in asymmetry energies of $\sigma=0.1\,T_{c_0}$. These parameters are close to those for H tunneling in Nb. Note that the Lorentzian and modified Wipf-Neumaier distributions have tails that decay as $1/E^2$.

\section{Effects of TLS impurities on superconductivity}\label{sec-Tc_Gap_Enhancement}

The thermodynamic properties of a superconductor with a concentration of TLS impurities depends on the order parameter obtained from a self-consistent solution of the gap equation Eq.~\eqref{eq-renormalized_gap_equation-regulated} and the renormalization factors, Eqs.~\eqref{eq-renormalization_functons},~\eqref{eq-Z_Delta} and~\eqref{eq-Z_varepsilon}, obtained from the inelastic contribution to the TLS-impurity self energy,~\eqref{eq-Self_Energy-inelastic}.
The effects of inelastic scattering by TLS impurities are determined by the distribution of tunnel splittings and the overall strength of the interaction between the TLS impurities and conduction electrons expressed in terms of the scattering rate $1/\tau_{in}=2\pi n_s N(0)\langle|m_{\vk,\vk'}|^2\rangle_{\Omega}$, where $n_s$ is mean density of TLS impurities, $N(0)$ is the normal-state quasiparticle density of states at the Fermi level and $\langle|m_{\vk,\vk'}|^2\rangle_{\Omega}$ is Fermi-surface average over the inelastic contribution to the quasiparticle-TLS scattering probability.
It is useful to express this scattering rate in units of the rate set by the timescale, $\tau_0=\hbar/2\pi T_{c_0}$, for Cooper pair formation in the clean limit, i.e. $\alpha=\tau_0/\tau_{in} = \hbar/2\pi\tau_{in} T_{c_0}$. Thus, $\alpha \ll 1$ corresponds to weak TLS impurity scattering, $\alpha\gg 1$ is the corresponding ``dirty limit'', and $\alpha=1$ is the intermediate disorder limit.

\subsubsection{Equilibrium Order Parameter and $T_c$ enhancement}

\begin{figure}
\centering
\includegraphics[width=0.75\textwidth]{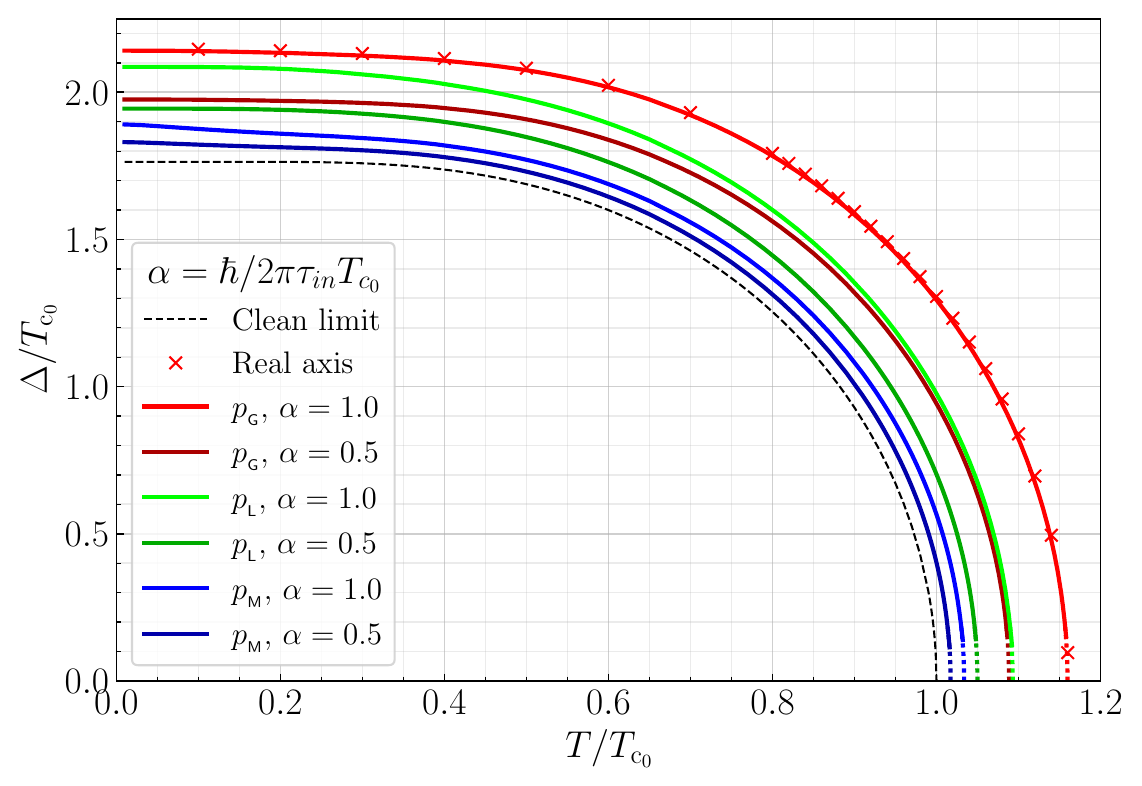}
\begin{minipage}{0.75\linewidth}
\caption{Enhancement of $T_c$ and the order parameter $\Delta$ versus temperature for the glass ($p_{_\mathsf{G}}$), Lorentzian ($p_{_\mathsf{L}}$) and modeifed Wipf-Neumaier ($p_{_\mathsf{M}}$) distributions of tunnel splittings, for inelastic scattering rates $\alpha\in\{0.5,\,1.0\}$. The \red{$\times$} symbols are solutions for $\Delta$ obtained from the gap equation
defined on the real energy axis, for the glass distribution and $\alpha=1.0$.
}
\label{fig-gap-vs-T}
\end{minipage}
\end{figure}

Figure~\ref{fig-gap-vs-T} shows results for the order parameter, $\Delta(T)$, obtained from self-consistent solutions of Eqs. \eqref{eq-renormalized_gap_equation-regulated}, \eqref{eq-renormalization_functons}, \eqref{eq-Z_Delta} and \eqref{eq-Z_varepsilon}, for the three models for the TLS distribution of tunnel splittings shown in Fig.~\ref{fig-tls-dist}, and for interaction strengths $\alpha=\hbar/2\pi\tau_{in}T_{c_0}$ spanning weak to intermediate disorder.

In all three models inelastic scattering by TLS impurities leads to enhancement of superconductivity, both an increase in $T_c$ as well as the order parameter (``gap'' ) at all temperatures. The gap enhancement is similar to that in superconductors with retardation from the electron-phonon interaction, c.f. Ref.~\cite{zar22} and references therein.
Indeed the largest enhancement of $T_c$ and $\Delta(T)$ is exhibited by glass distribution which has the broadest distribution of tunnel splittings,
and thus the largest relative ground state occupancy, $N_{ge}$. 
By contrast the modified Wipf-Neumaier distribution is peaked at $E=T_{c_0}$, and thus has a significant population of TLS impurities in excited states.
Distributions dominated by TLS level splittings with $E\ge T_{c_0}$ give rise to larger gap enhancement. 

For equilibrium TLS populations these impurities are predominantly in their ground state, and effectively static impurities. Thus, only the high energy tail of the distribution contributes to the gap enhancement.

The order parameter can also be calculated by analytically continuing Eq.~\eqref{eq-renormalized_gap_equation-regulated} and the self energies to real energies. For TLS impurities in equilibrium with the electrons and lattice, the two formulations are equivalent and must give the same result for $\Delta(T)$.
The comparison of the solutions based on the Matsubara gap equation and the real-frequency forumulation are shown for the glass distribution and $\alpha=1.0$; the results shown as the cross marks are in exact agreement with the solution obtained from the Matsubara gap equation.

\subsubsection{Relation to earlier research on TLS impurities in superconductors}\label{sec-earlier_TLS-QP_research}

Early on several authors showed that $T_c$ could be enhanced by impurities with discret level splittings~\cite{ful70,bra70,vuj79,mae84}. 
Fulde, Hirst and Luther considered superconductors containing paramagnetic rare earth impurities with crystal-field split energy levels~\cite{ful70}. The latter leads to enhancement of $T_c$ similar to that arising from optical phonons, and as we discussed inelastic scattering by TLS impurties. The enhancement competes with pair-breaking from spin-exchange scattering of conducting electrons by the paramagnetic impurity.    
Brandt also considered inelastic scattering by impurties with a level splitting, and obtained a result for the enhancement of $T_c$ to first order in the impurity concentration, as a function of $E/T_{c_0}$.

Vujičić et al.~\cite{vuj79} considered TLS impurities embedded in the lattice of high-T$_c$ superconductors, with each TLS impurity localized on the sites of an impurity lattice. Each TLS impurity was assumed to have the same tunnel splitting. In this model the localized TLS impurities are highly correlated in space, with long-range interactions. The authors then related inelastic electron-TLS scattering to an Eliashberg function, $\alpha^2(\omega)F(\omega)_{tls}$, and assumed the electron-TLS coupling to be the same magnitude as that for the coupling of band electrons to phonons of the pure superconductor. This is essentially an Einstein phonon spectrum except for dispersion of the TLS energy splittings spread over a narrow bandwidth, $\omega_{tls}\ll\omega_{D}$, compared to the phonon bandwidth $\omega_D$. In the limit $T_c \ll \omega_{tls}\ll \omega_D$ the authors obtained the result
\begin{equation}
T_c = 1.14\,\omega_{tls}\exp{-1/\lambda_{\mathrm{eff}}}
\,\,\mbox{with}\,\,
\lambda_{\mathrm{eff}} = \lambda_{tls} + 
\frac{\lambda_{ph}-\mu^*}{1-(\lambda_{ph}-\mu^*)\ln(\omega_D/\omega_{tls})}
\,,
\end{equation}
where $\lambda_{a}=\int\,d\omega\,\alpha^2F(\omega)_a/\omega$ is the coupling constant for TLS or phonon mediated pairing, i.e. $a\in\{tls,ph\}$, and 
$\mu^*$ is the Coulomb pseudo-potential.
The authors argue that for $\omega_{tls}\ll\omega_D$ and for nearly complete occupation of TLS impurities in the ground level, $N_g\gg N_e$, that $\lambda_{tls}\gg\lambda_{ph}$. This leads to extremely large enhancement of the transition temperature, $T_c/T_{c_0}\simeq 27$, relative to that predicted by the electron-phonon interaction. 
The model and results differ dramatically from our predictions even for TLS impurities described by a sharp distribution of tunnel splittings. Given the model of a lattice of identical TLS impurities, and assumptions for the strength of the electron-TLS coupling, the overlap of this work with our theory is only the qualitative prediction that TLS impurities lead to an enhancement of $T_c$.

The formulation of the interaction of embedded TLS impurities in a superconductor by Maekawa et al.~\cite{mae84}, based on tunneling impurities described by an $\point{SU(2)}{\ns}$ isospin, is closest to our formalism. However, these authors were interested in the Kondo limit in which the tunnel splitting $E\rightarrow 0^+$. In this limit the inelastic processes we consider are absent. Their results based on higher order scattering processes lead to weak pairbreaking and the suppression of $T_c$ in the limit $E\ll T_{c_0}$.

A main result of our microscopic theory for the interaction of a random distribution of TLS impurities on superconductivity, for a broad range to distributions of tunnel splittings, is that $T_c$ and gap enhancement are determined by the high-energy part of the spectrum of TLS impurites, i.e. TLS impurities with $E\gtrsim T_{c_0}$. 

\subsubsection{TLS induced superconductivity}

The enhancement of the superconductivity by the random distribution of TLS impurities suggests that the kernel of the off-diagonal self energy, $\Sigma_{\Delta}$, resulting from inelastic scattering by TLS impurities \emph{generates} a retarded pairing interaction and corresponding TLS generated gap equation.~\footnote{Riess and Maynard considered the possibility of pairing induced by TLS impurities in metallic alloys~\cite{rie80}. These authors adopted the amorphous glass distribution of TLS levels and obtained a formula analogous to McMillan's formula for $T_c$ as a function of the electron-phonon interaction, but with $\lambda_0$ replaced by $\sqrt{\lambda_0}$ for TLS-mediated pairing. Our analysis is in the opposite limit and 
makes no approximations to the frequency dependenced of the TLS-induced paring interaction.}
The latter is obtained from Eq.\eqref{eq-renormalization_functons} with $\Delta=0$, 
\begin{eqnarray}
\tilde\Delta(\varepsilon_n)
&=& 
\frac{1}{2\pi\tau_{in}}\,
\int dE\,p(E)\,N_{ge}(E)\,\pi T\sum_{\varepsilon_{n'}}\,
\mfD(\varepsilon_n-\varepsilon_{n'};E)\,
\frac{\tilde\Delta(\varepsilon_{n'})}
     {\sqrt{(\varepsilon_{n'}+i\Sigma_{\epsilon}(\varepsilon_{n'})^2
     +\tilde\Delta(\varepsilon_{n'})^2}}
\,,
\label{eq-tildeDelta}
\end{eqnarray}
where $\tilde\varepsilon_n=\varepsilon_n+i\Sigma_{\epsilon}(\varepsilon_n)$, 
with
\begin{eqnarray}
i\Sigma_{\epsilon}(\varepsilon_n)
&=& 
\frac{1}{2\pi\tau_{in}}\,
\int dE\,p(E)\,N_{ge}(E)\,\pi T\sum_{\varepsilon_{n'}}\,
\mfD(\varepsilon_n-\varepsilon_{n'};E)\,
\frac{\varepsilon_{n'} + i\Sigma_{\epsilon}(\varepsilon_{n'})}
     {\sqrt{(\varepsilon_{n'}+i\Sigma_{\epsilon}(\varepsilon_{n'})^2
     +\tilde\Delta(\varepsilon_{n'})^2}}
\,.
\label{eq-tildevarepsilon}
\end{eqnarray}

\begin{figure}
\centering
\includegraphics[width=0.65\textwidth]{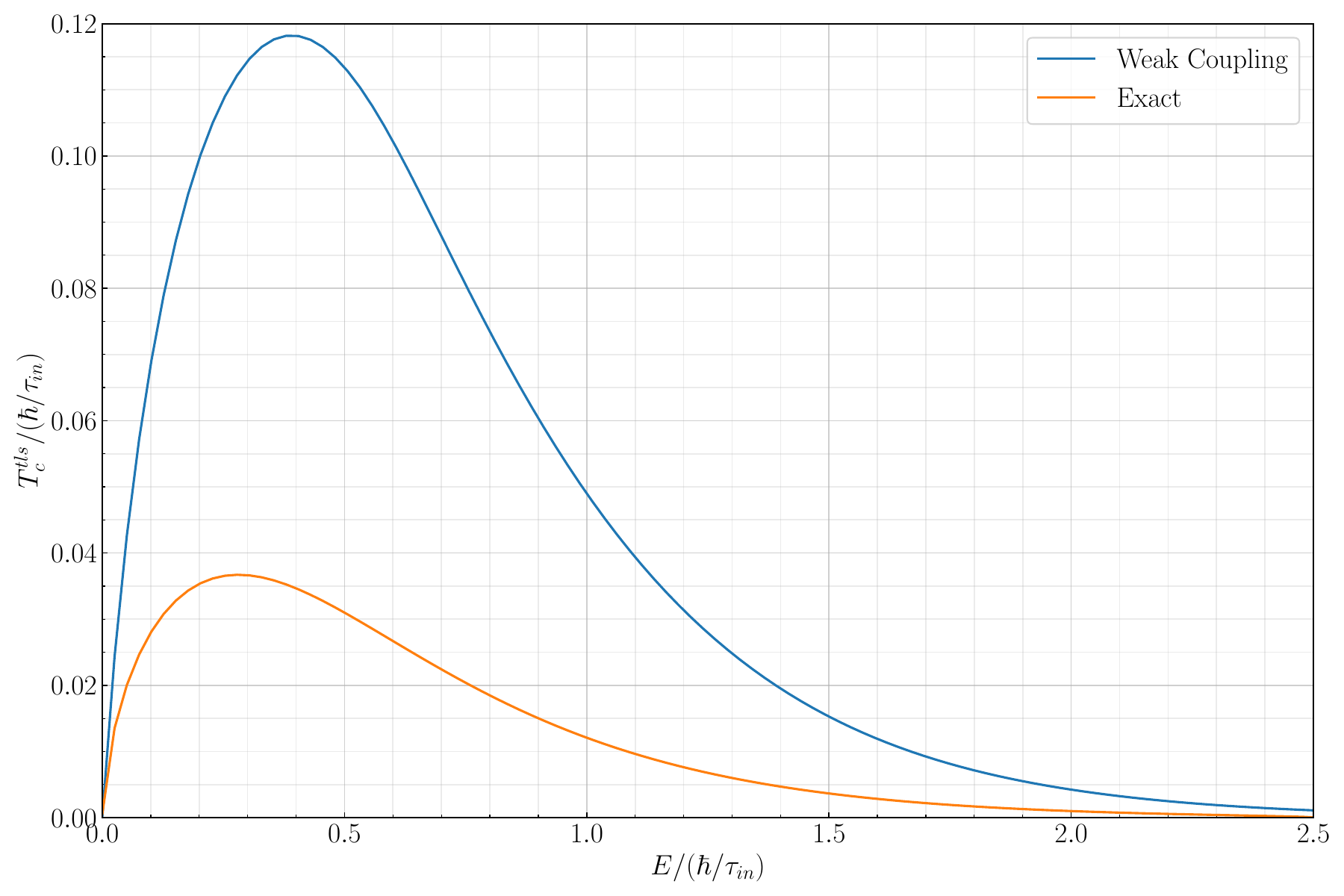}
\begin{minipage}{0.75\linewidth}
\caption{
Superconducting transition temperature, $T_c$, induced by electron-TLS scattering versus tunnel splitting, $E$. Both $T_c$ and the tunnel splitting are scaled in units of the energy associated with the inelastic scattering rate, $\hbar/\tau_{in}$. The ``Weak Coupling'' result based on Eq.~\eqref{eq-Tc-TLS_weak-coupling} substantially overestimates $T_c$ for electron-TLS mediated superconductivity.
}
\label{fig-TLS-induced-Tc}
\end{minipage}
\end{figure}

The onset of superconductivity for electron-TLS mediated pairing is obtained by linearizing Eq.~\eqref{eq-tildeDelta} in $\tilde\Delta(\varepsilon_n)$, and setting $\tilde\Delta(\varepsilon_n)=0$ in Eq.~\eqref{eq-tildevarepsilon} for $\Sigma_{\epsilon}$. The resulting homogeneous equation for $\tilde\Delta$ is 
\begin{eqnarray}
\tilde\Delta(\varepsilon_n)
&=& 
\frac{1}{2\pi\tau_{in}}\,
\int dE\,p(E)\,N_{ge}(E)\,\pi T\sum_{\varepsilon_{n'}}\,
\mfD(\varepsilon_n-\varepsilon_{n'};E)\,
\frac{\tilde\Delta(\varepsilon_{n'})}
     {\vert\varepsilon_{n'}+i\Sigma_{\epsilon}(\varepsilon_{n'})\vert}
\,,
\label{eq-tildeDelta-linear}
\end{eqnarray}
where $\Sigma_{\epsilon}$ reduces to
\begin{eqnarray}
i\Sigma_{\epsilon}(\varepsilon_n)
&=& 
\frac{1}{2\pi\tau_{in}}\,
\int dE\,p(E)\,N_{ge}(E)\,\pi T\sum_{\varepsilon_{n'}}\,
\mfD(\varepsilon_n-\varepsilon_{n'};E)\,\sgn{\varepsilon_{n'}}
\,.
\label{eq-tildevarepsilon-linear}
\end{eqnarray}
Equation~\eqref{eq-tildeDelta-linear} can be expressed as a matrix eigenvalue equation in the space of Matsubara frequencies with indices $(n,n')$, 
\begin{equation}
\sum_{n'}\left(K_{n,n'} - \delta_{n,n'}\right)\,
\tilde\Delta(\varepsilon_{n'})
= 0
\,, 
\end{equation}
where 
\begin{equation}
K_{n,n'} 
= \frac{\hbar}{2\pi\tau_{in}}\,
\int dE'\,p(E')\,N_{ge}(E')\,
\pi T\,
\mfD(\varepsilon_n-\varepsilon_{n'};E')\,
\frac{1}{\vert\varepsilon_{n'}+i\Sigma_{\epsilon}(\varepsilon_{n'})\vert}
\Bigg\vert_{T_c^{tls}}
\,.
\label{eq-Kernel_linearized_TLS-gap_equation}
\end{equation}
is the kernel of the pairing self energy.
If a non-trivial solution exists with $|\tilde\Delta(\varepsilon_n)|\xrightarrow{T\rightarrow T_c^{tls}-0^+} 0^+$ for $T_c^{tls}>0$, then the superconducting transition temperature is determined by the vanishing determinant, 
\begin{equation}
\mathsf{Det}\left(K_{n,n'} - \delta_{n,n'}\right)\Big\vert_{T_c^{tls}}=0
\,.
\label{eq-Determinant}
\end{equation}

For TLS impurities with a delta function distribution, $p(E') = \delta(E'-E)$,
i.e. all TLS impurities with the same tunnel splitting $E$, Eq.~\eqref{eq-Kernel_linearized_TLS-gap_equation} simplifies to
\begin{equation}
K_{n,n'} 
= \frac{\hbar}{\tau_{in}}\,N_{ge}(E)\,T_c^{tls}\,
\frac{E}{E^2+(\varepsilon_n-\varepsilon_{n'})^2}
\frac{1}{\vert\varepsilon_{n'}+i\Sigma_{\epsilon}(\varepsilon_{n'})\vert}
\Bigg\vert_{T_c^{tls}}
\,.
\label{eq-Kernel_linearized_TLS-gap_equation-fixed-E}
\end{equation}
The kernel decays rapidly for $|\varepsilon_n|,\,|\varepsilon_{n'}|\gg E$.
Thus, we introduce a cutoff $|\varepsilon_{n,n'}|\le E_c$ with $E_c \ge E$, then evaluate the determinant in Eq.~\eqref{eq-Determinant} as a function of $\hbar/\tau_{in}$ and $E$, and locate the zeroes.

In the limit $E\gg 2\pi T_c^{tls}$ we replace the Bosonic propagator by
$E/(E^2+(\varepsilon_n-\varepsilon_{n'})^2)\rightarrow 1/E$ for 
$|\varepsilon_n|,|\varepsilon_{n'}|\le E$, which is analogous to the weak-coupling limit for phonon-mediated superconductivity, but with $E$ analogous to the maximum phonon frequency. In this limit 
$\tilde\Delta(\varepsilon_n) = \Delta \Theta(E-|\varepsilon_n|)$, in which case $i\Sigma_{\epsilon} = 0$, and thus the linearized gap equation becomes,
\begin{equation}
1=\frac{\hbar}{\tau_{in}E}\,N_{ge}(E)\,T_c^{tls}\,
\sum_{n}^{|\varepsilon_n|\le E}\,
\frac{1}{|\varepsilon_n|}\Big\vert_{T_c^{tls}}
\,.
\end{equation}
Evaluating the diGamma function (see paragraph after Eq.~\eqref{eq-gap_equation_clean-limit}) in the limit $E\gg T_c^{tls}$ gives the transcendental equation for $T_c^{tls}$,
\begin{equation}
T_c^{tls}
=
\frac{2e^\gamma}{\pi}\,E\,\exp{-\frac{\pi E}{N_{ge}(E)\,\hbar/\tau_{in}}}
\approx
1.13\,E\,\exp{-\frac{\pi E}{N_{ge}(E)\,\hbar/\tau_{in}}}
\,.
\label{eq-Tc-TLS_weak-coupling}
\end{equation}

\begin{figure}
\begin{center}
\includegraphics[width=0.4\textwidth]{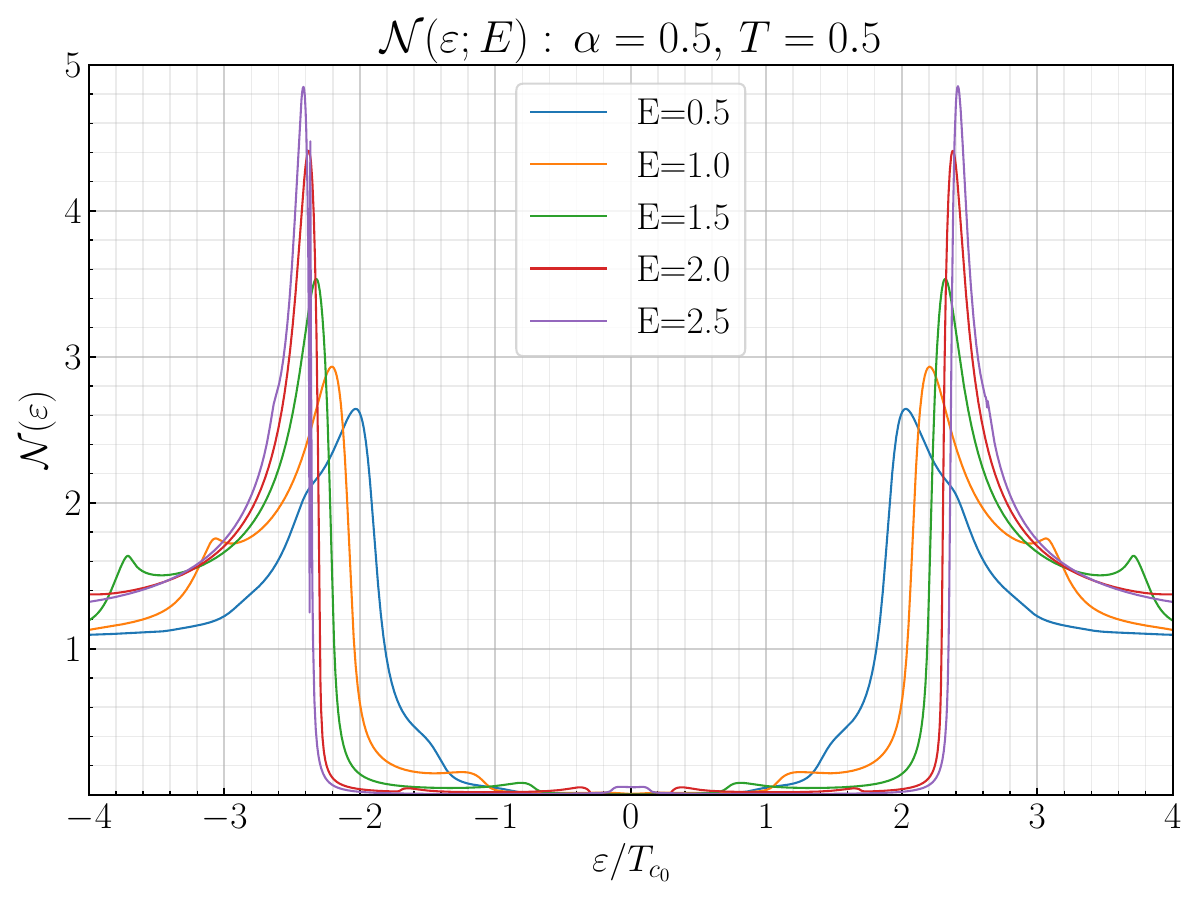}~\includegraphics[width=0.4\textwidth]{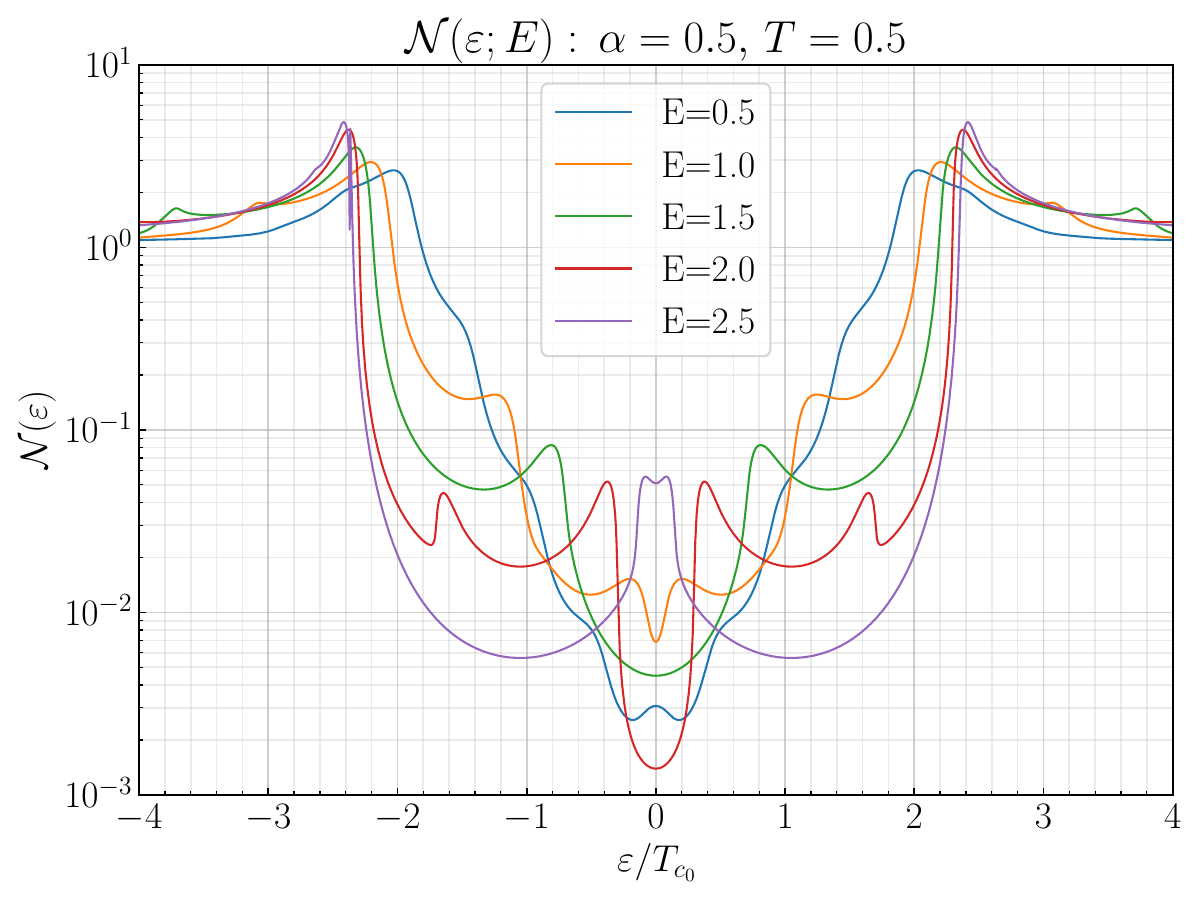}
\includegraphics[width=0.4\textwidth]{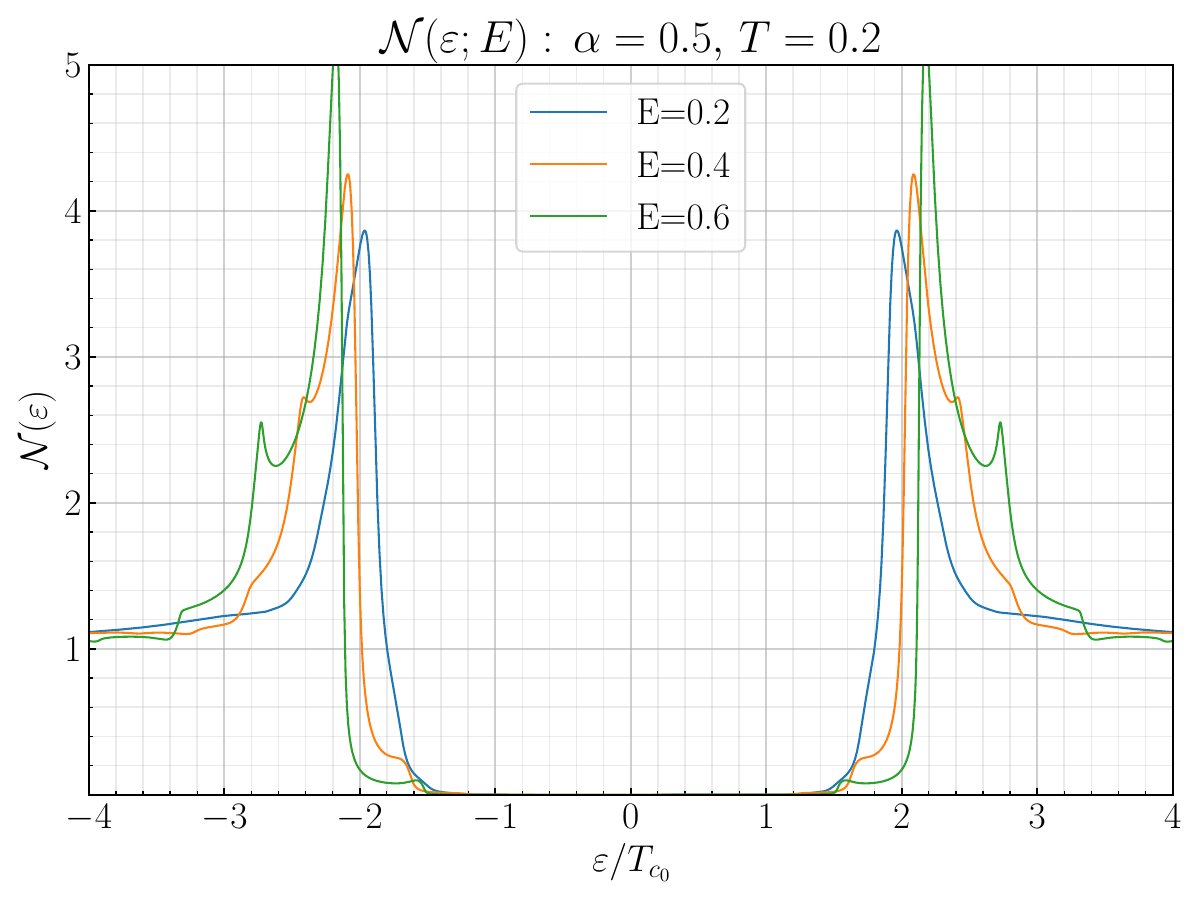}~\includegraphics[width=0.4\textwidth]{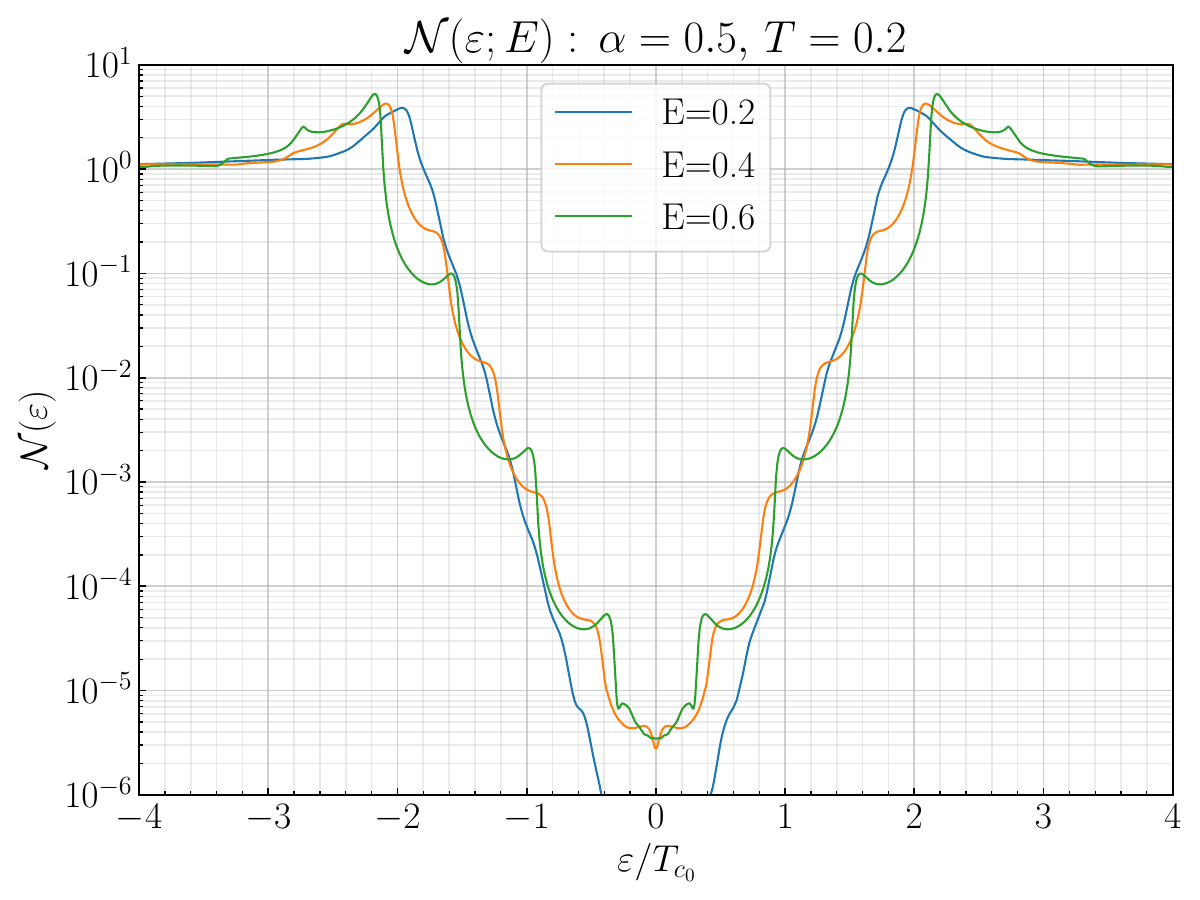}
\end{center}
\begin{minipage}{0.8\textwidth}
\caption{The quasiparticle DOS for several delta function distributions. 
The top row is the equilibrium DOS for $T=0.5\,T_{c_0}$, 
linear and log scales, showing sub-gap quasiparticle states and above gap resonances. The bottom row corresponds to $T=0.2\,T_{c_0}$, and illustrates the 
sensitivity of the subgap DOS to level population of the TLS impurities. 
}
\label{fig-dos-dist}
\end{minipage}
\end{figure}

Note that the difference in populations of the ground and excited states enters the exponent, and as a result the solution for $T_c^{tls}$ is very sensitive to the relative level populations. 
For $N_{ge}>0$ there is always a superconducting transition. However, there is no physical solution for $N_{ge}\le 0$.
The weak-coupling formula, Eq.~\eqref{eq-Tc-TLS_weak-coupling}, for $T_c^{tls}$ is plotted in Fig.~\ref{fig-TLS-induced-Tc}.
Note that at the peak value of $T_c^{tls}$ we have $E/2\pi T_c^{tls}\approx 0.5$, in which case the assumption underlying weak-coupling approximation is far from being satisfied. 
This is born out by numerical solutions of Eq.~\eqref{eq-Determinant} with the kernel in Eq.~\eqref{eq-Kernel_linearized_TLS-gap_equation-fixed-E} with the results also shown in Fig.~\ref{fig-TLS-induced-Tc}. The numerical determination of $T_c^{tls}$ is obtained by choosing a cutoff energy $E_c \gg \mbox{max} 
\{2\pi T,E\}$ that defines the maximum Matsubara energy, or $N_c=\mbox{Int}(E_c/2\pi T)$, i.e. the dimension of matrix $K_{n,n'}$. 
We find that for $N_c=10^3$ the numerical result for $T_c^{tls}$ is converged and insensitive to extending the dimension of the matrix $K_{n,n'}$.

It is interesting to consider the possibility that TLS disorder introduced into clean non-superconducting metals such as Au, Cu, Pt etc. might lead to superconductivity, but it is also important to note that repulsion from the Coulomb pseudo potential may compensate the attractive interaction mediated by TLS impurities.

\section{Sub-gap states, pair-breaking \& nonequilibrium TLS level populations}\label{sec-Nonequilibrium_TLS}

Solving the real-frequency gap equation and retarded self energies is computationally expensive compared to the Matsubara formulation. However, it has the advantage that the spectrum of electronic states contributing to the order parameter and the self energy are obtained explicitly, and thus separate from the distributions of occupied and unoccupied quasiparticle energies and TLS levels.

\subsection{Density of states}

In particular, the quasiparticle density of states, $\cN(\varepsilon)$, for $T<T_c$ is computed from Eqs.~\eqref{eq-DOS}, and~\eqref{eq-Propagator-retarded}, once we have calculated the order parameter and the retarded self energies ~\eqref{eq-Self_energy-retarded}. 
Inelastic scattering of quasiparticles that are bound to form Cooper pairs leads both 
pair breaking in addition to enhancement of superconductivity. 
For the TLS impurities with level populations in equilibrium with the thermal bath, pair breaking is weaker than pair enhancement. Nevertheless, subgap quasiparticle states indicative of pair breaking appear in the quasiparticle density of states as shown in Fig.~\ref{fig-dos-dist}. The first row corresponding to $T=0.5T_{c_0}$ shows subgap states at all energies down to the Fermi level ($\varepsilon=0$), with $\cN(0)/N_f\sim 0.01-0.1$, as well as above-gap resonances, particulary for the largest tunnel splitting, $E=2.5\,T_{c_0}$. The spectral weight of the sub-gap states is suppressed at lower temperatures, e.g. $T=0.2\,T_{c_0}$ shown in the second row, as the TLS impurities are more likely to be in the ground energy level.

\begin{figure}
\centering
\includegraphics[width=0.45\textwidth]{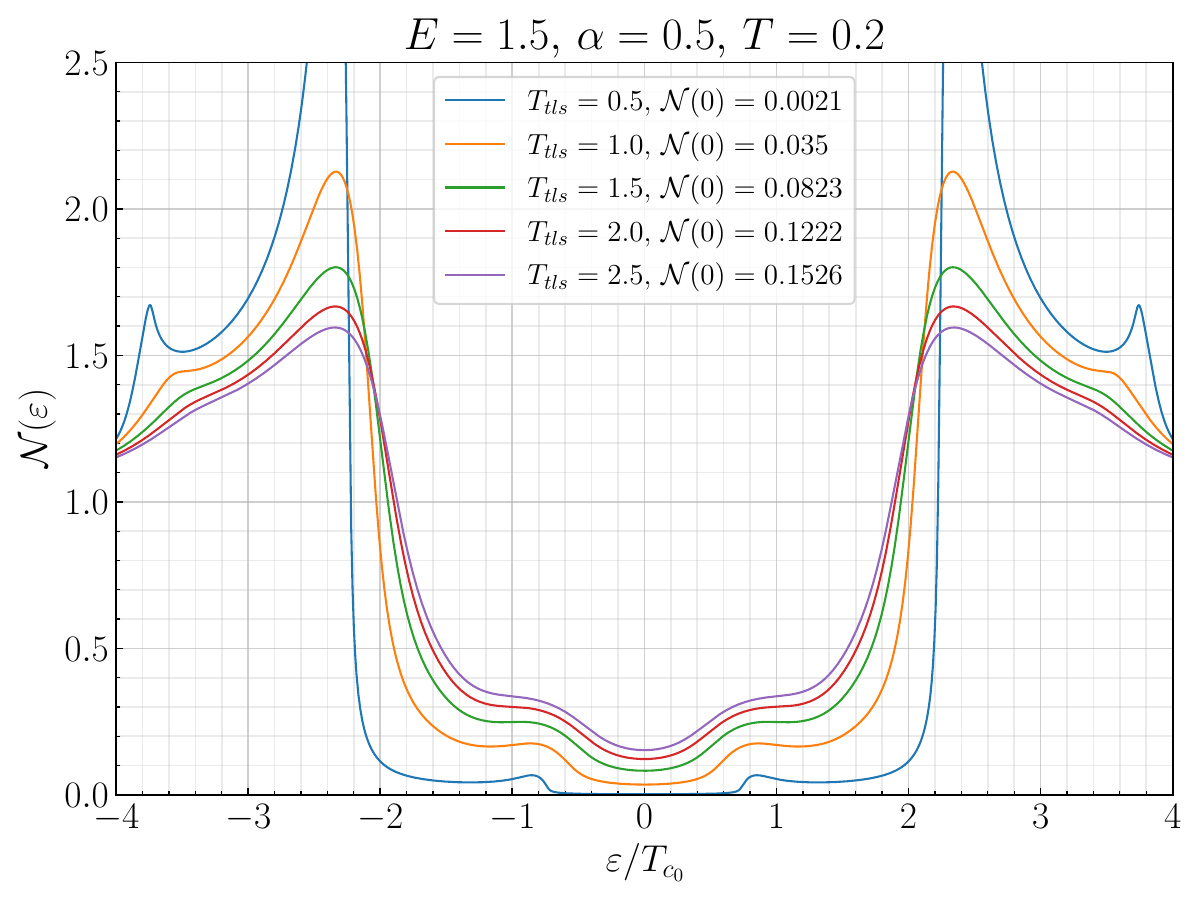}~\includegraphics[width=0.45\textwidth]{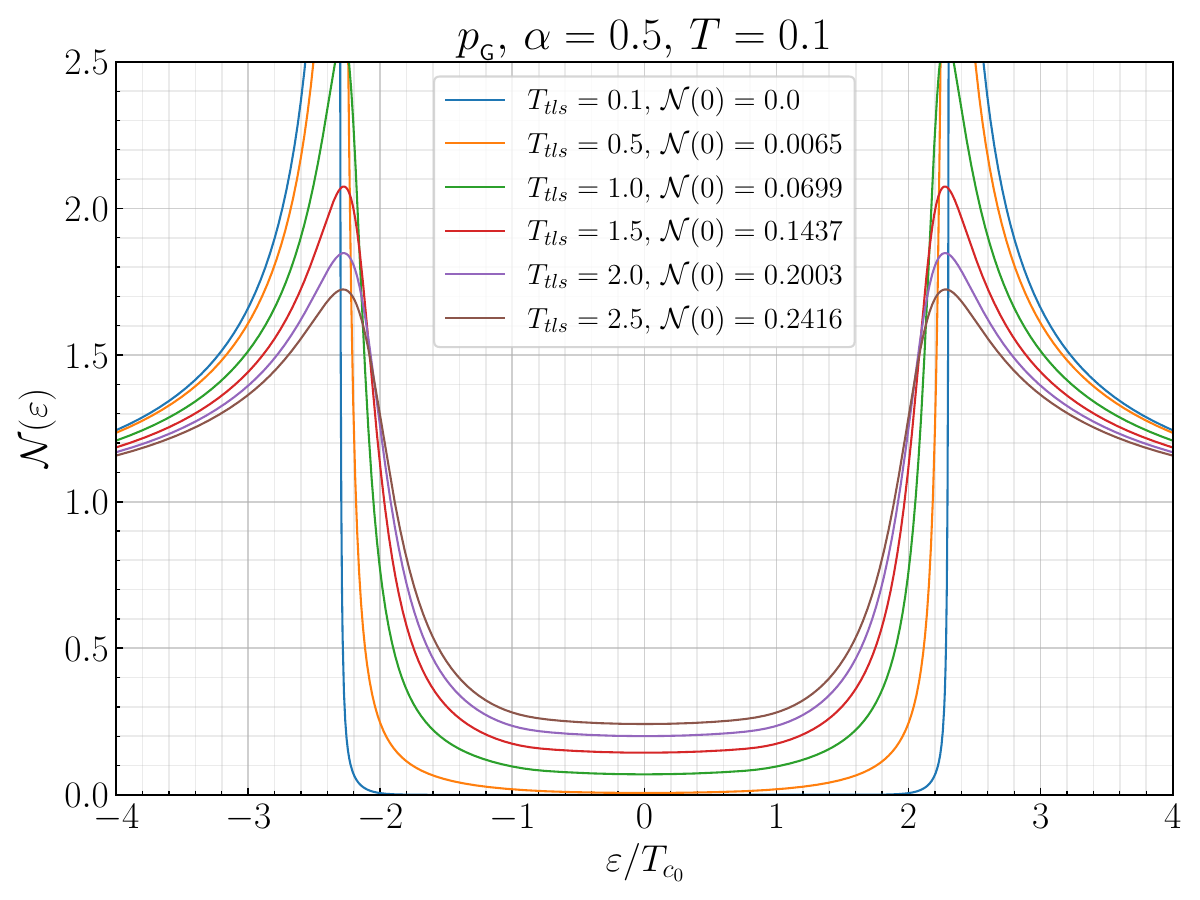}
\includegraphics[width=0.45\textwidth]{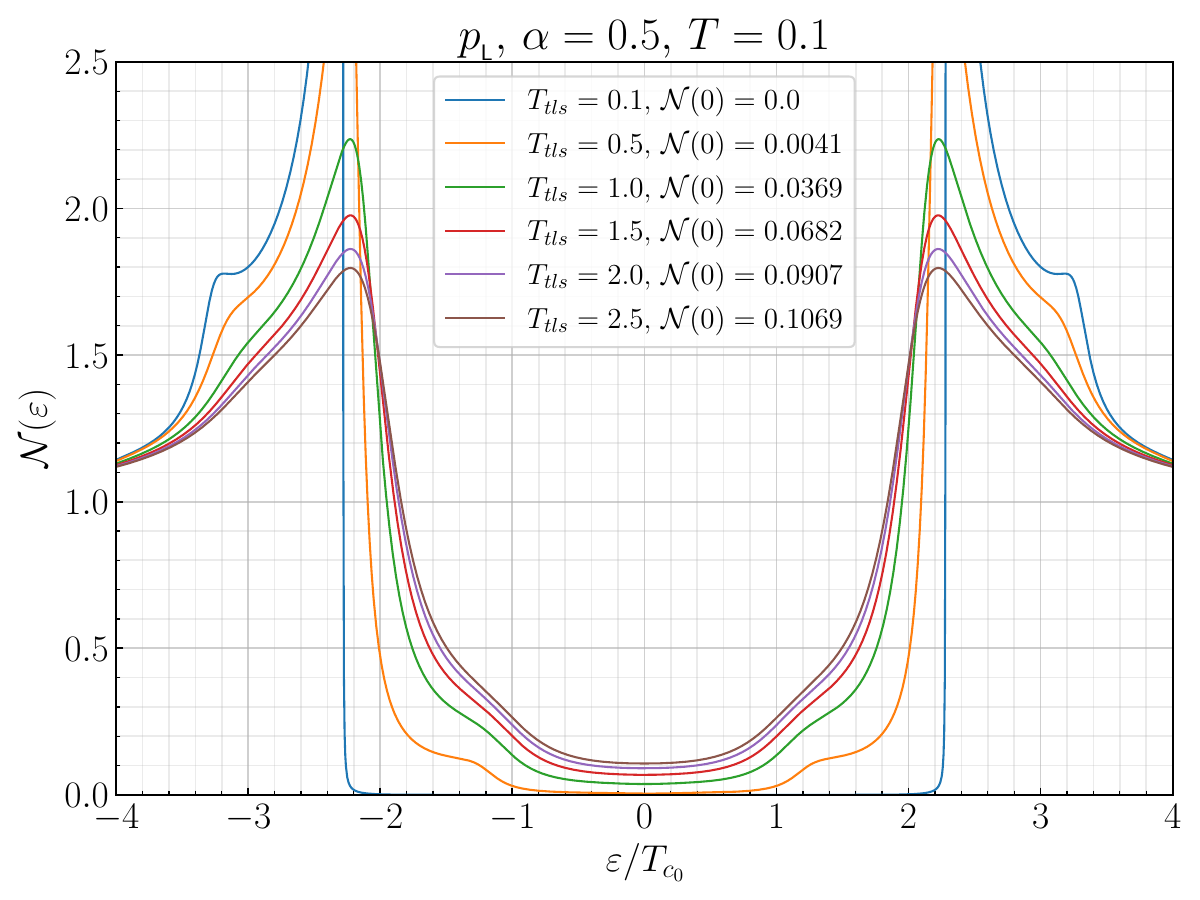}~\includegraphics[width=0.45\textwidth]{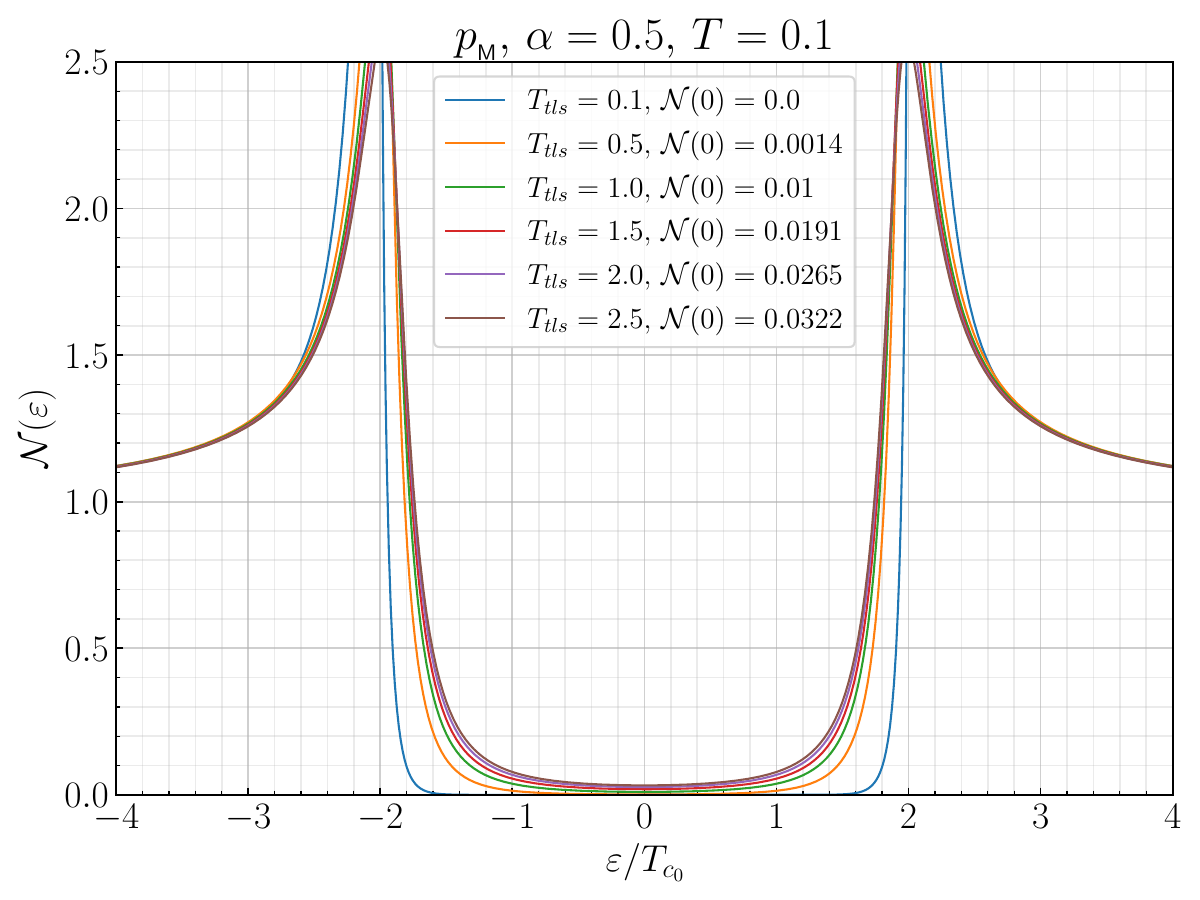}
\begin{minipage}{0.9\textwidth}
\caption{
The quasiparticle DOS for four TLS distributions: delta function (top left), glass-like (top right), Lorentzian (bottom left) and modified Wipf-Neumaier (bottom right). The level populations of the TLS impurities are out of equilibrium with the thermal bath, and described by the effective temperature $T_{tls}$.
Pair-breaking by the non-equilibrium TLS population leads to a gapless spectrum for all TLS models.
}
\label{fig-dos-noneq}
\end{minipage}
\end{figure}
\subsubsection*{Enhanced Pair-Breaking by Nonequilibrium TLS Level Populations}

The sensitivity of the sub-gap spectrum of quasiparticles and pairing self energy to the relative populations of the excited and ground states of the distribution of TLS impurities also implies that ineffective equilibration of the TLS population to the conduction electrons bath and the lattice can have substantial impact on the quasiparticle spectrum, and thus the response of such supercondcutors to microwave and acoustic fields. 
Indeed recent experiments using immersion cooling of superconducting qubits and resonators in liquid \He\ show that the TLS population is relaxed substantially to their ground levels~\cite{luc23}. In particular, the microwave power required to saturate the TLS levels increases by three orders of magnitude with immersion cooling, suggesting that the TLS populations may be substantially out of equilibrium for superconducting devices operating under microwave or current excitation.
  
To exhibit the sensitivity of the spectrum of sub-gap quasiparticle states to the TLS level population start from the specral function of the TLS impurity self energy,
\begin{eqnarray}
\Im\widehat{\Sigma}^{\mathrm{R}}(\varepsilon) 
=
\frac{\hbar}{2\pi\tau_{in}}\int dE\,p(E)
&\ns\Big\{&
\ns
\Im\whmfG^{\mathrm{R}}(\varepsilon-E)\,
\left[\cdbox{N_g(E)-N_{ge}(E)}\,\nf(\varepsilon-E)\right]
\nonumber\\
&+&
\Im\whmfG^{\mathrm{R}}(\varepsilon+E)\,
\left[\cdbox{N_e(E)+N_{ge}(E)}\,\nf(\varepsilon+E)\right]
\Big\}
\,,
\label{eq-im-SE}
\end{eqnarray}
where the dependence on the occupation of the ground and excited levels is highlighted.  
Multiple factors can lead to TLS levels being out of equilibrium with the bath of electrons and phonons, including the relaxation timescales of the TLS levels via electron and phonon interactions, microwave fields coupling to TLS impurities, and screening currents driving the TLS distributions.
Here we model the out of equilibrium TLS level population by an effective temperature of the TLS level population, $T<T_{tls}\le 2.5\,T_{c_0}$. 
The nonequilibrium TLS population dramatically changes the spectrum of sub-gap states, particularly near the Fermi energy ($\varepsilon=0$), providing a mechanism for pair-breaking that spans all energies below the clean limit excitation gap, in a way that is remarkably similar to the ad hoc Dynes pair-breaking parameter~\cite{lec20}.

Figure~\ref{fig-dos-noneq} shows the impact of the nonequilibrium level populaton of TLS impurities parametrized by the effective TLS bath temperature on the quasiparticle density of states for four TLS level distributions, and with electrons and phonons at $T=0.1-0.2 T_{c0}$.
The primary observation is that the sub-gap density of states at the Fermi level is exponentially sensitive to the TLS level population that is out of equilibrium. The quasiparticle DOS is also sensitive to the distribution of TLS energy level splittings. In particular, scattering off tunneling impurities with distributions that have a substantial density of high-energy level splittings generate the largest DOS at the Fermi level. Thus, the glass-like distribution with $E_{c}=3.0\,T_{c_0}$ exhibits the largest DOS at the Fermi level, while the modified Wipf-Neumaier distribution, which is dominated by TLS impurities with much lower-energy tunnel splittings, exhibits a finite, but significantly lower DOS at the Fermi level.   

The significance of the sub-gap quasiparticle spectrum generated by out of equilibrium TLS impurities is the {\it new channel for microwave power loss} from quasiparticle dissipation that results from pair-breaking by scattering from non-equilibrium TLS impurities.
A quantitative analysis of quasiparticle dissipation generated by the non-equilibrium TLS population is the subject of a separate study that includes the population dynamics of the TLS impurity distribution driven by a microwave field.

\section{Resum\`e}

We have developed a quantum field theory formulation of superconductors interacting with a random distribution of TLS impurities. The basic interaction is formulated as a scattering theory of quasiparticles, including those bound as Cooper pairs, and TLS impurities, with the internal state of each impurity described by an $\point{SU(2)}{\ns}$ isospin.
Our formulation represents each TLS isospin by a local fermion field \`a la Abrikosov~\cite{abr65}, but projects out unphysical states based on Popov and Fedetov's method~\cite{pop88} that ensures the resulting perturbation theory in the quasiparticle-TLS interaction obeys the Wick theorem.
The most important scattering processes are those in which incoming and outgoing quasiparticles exchange energy as a TLS impurity undergoes a transition between its ground and excited state. 
This inelastic process leads to both pair-enhancement and pair-breaking. 

Pair-enhancement dominates at low temperatures under equilibrium conditions of quasiparticles, phonons and TLS impurities, and leads to modest increase of $T_c$ and the superconducting order parameter, $\Delta$. 
Pair breaking is evident from sub-gap quasiparticle states, even for TLS impurities in equilibrium with the thermal bath.
A key result of our analysis is the sensitivity of the sub-gap density of states to the level populations of the distribution of TLS impurities.
Thus, under excitation, or decoupling from the thermal bath, the resulting nonequilibrium level population of the TLS distribution generates subgap quasiparticle states down to the Fermi level which contribute to dissipation and thus degrade the performance of superconducting devices.

\begin{acknowledgments}
We thank Anna Grassellino, Jens Koch, Corey Rae McRae, Alex Romanenko and David Seidman for discussions on the role of TLS defects in Nb-based resonators and devices. 
This work was supported by the U.S. Department of Energy, Office of Science, National Quantum Information Science Research Centers, Superconducting Quantum Materials and Systems Center (SQMS) under contract number DE-AC02-07CH11359.
\end{acknowledgments}

%
\end{document}